\documentclass[letterpaper, 10 pt, conference]{ieeeconf}  

\IEEEoverridecommandlockouts                              
\usepackage{graphicx,subfigure} 
\usepackage{epsfig} 
\usepackage{amsmath} 
\usepackage{amssymb}  
\usepackage{ulem}

\title{\LARGE \bf Noise, Bifurcations, and Modeling of Interacting Particle Systems
}

\author{Luis Mier-y-Teran-Romero, Eric Forgoston and Ira B. Schwartz 
\thanks{This work is supported by the Office of Naval Research}
\thanks{Luis Mier-y-Teran-Romero is a joint NIH  postdoctoral fellow with the
   Johns Hopkins University and the US Naval Research Laboratory, Nonlinear Systems
   Dynamics Section, Code 6792, Washington, DC 20375, USA, 
        {\tt\small luis@nlschaos.nrl.navy.mil}}%
\thanks{Eric Forgoston is with the Department of Mathematical Sciences, Montclair State University,
        Montclair, NJ 07043, USA,
        {\tt\small eric.forgoston@montclair.edu}}%
\thanks{Ira B. Schwartz is with the Nonlinear Systems Dynamics Section, Code 6792, US
  Naval Research Laboratory,
        Washington, DC 20375, USA,
        {\tt\small ira.schwartz@nrl.navy.mil}}%
}

\begin{document}

\maketitle
\thispagestyle{empty}
\pagestyle{empty}

\begin{abstract}
We consider the  stochastic patterns of a system of communicating, or coupled, self-propelled particles
in the presence of noise and communication time delay. For sufficiently large
environmental noise, there exists a transition between a translating state and
a rotating state with stationary center of mass. Time delayed communication
creates a bifurcation pattern dependent on the coupling amplitude
between particles. Using a mean field  model in the large number limit, we show how the
complete bifurcation unfolds in the presence of communication delay and
coupling amplitude. Relative to the center of mass, the  patterns can then be described as transitions between
translation, rotation about a stationary point, or a rotating swarm, where the
center of mass undergoes a Hopf bifurcation from steady state to a limit
cycle. Examples of some of the  stochastic patterns will be given for large
numbers of particles.

\end{abstract}

\section{INTRODUCTION}

The collective motion of interacting multi-particle systems has been the subject of many
recent experimental and modeling studies. It is especially astounding that numerous coherent states of
great complexity can arise spontaneously in spite of the absence of a
particle acting as a leader. The study of these swarming systems has proven useful in
understanding the spatio-temporal patterns formed by bacterial colonies, fish,
birds, locusts, ants, pedestrians, etc.~\cite{Budrene95, Toner95, Toner98,
  Parrish99,EdelsteinKeshet98,Topaz04}. Moreover, these studies have provided
valuable information that may be exploited in the design of systems of
autonomous, inter-communicating robotic systems~\cite{Leonard02,Justh04,
  Morgan05}.

Investigators have used various mathematical approaches to study swarm
systems. Some studies have preserved the individual character of each agent in
the system, using ordinary or delay differential equations (ODEs/DDEs) to
describe their trajectories~\cite{Vicsek95,Flierl99,Couzin02,Justh04}. Other researchers have proposed continuum models written in
terms of averaged velocity and particle density fields
that satisfy partial differential equations (PDEs)~\cite{Toner95,Toner98,EdelsteinKeshet98, Topaz04}. In addition, authors have
also studied the effects of noise in the swarms and shown the existence of
noise-induced transitions~\cite{Erdmann05, Forgoston08}.

More recently, authors have begun to study the effects of communication
time-delays between particles. Time-delay models are common in many areas of
mathematical biology including population dynamics, neural
networks, blood cell maturation, virus
dynamics and genetic networks~\cite{MacDonald78, MacDonald89,Campbell02,
  Bernard04, Mackey04a, Mackey04b, TianJensSnepp02, JensSnepp03, Monk03}. In
the context of swarming particles, it has been shown that the introduction of a
communication time-delay may induce transitions between different coherent
states~\cite{Forgoston08}. The type of transition is dependent on the coupling
strength between particles and the noise intensity.

Here we make a more detailed study of the bifurcation structure of the mean
field approximation to the delay-coupled model
proposed studied in~\cite{Forgoston08} and investigate how the
bifurcations in the system are modified in the presence of noise.

\begin{figure}[h]
\begin{minipage}{0.49\linewidth}
\includegraphics[width=4.3cm,height=3.0cm]{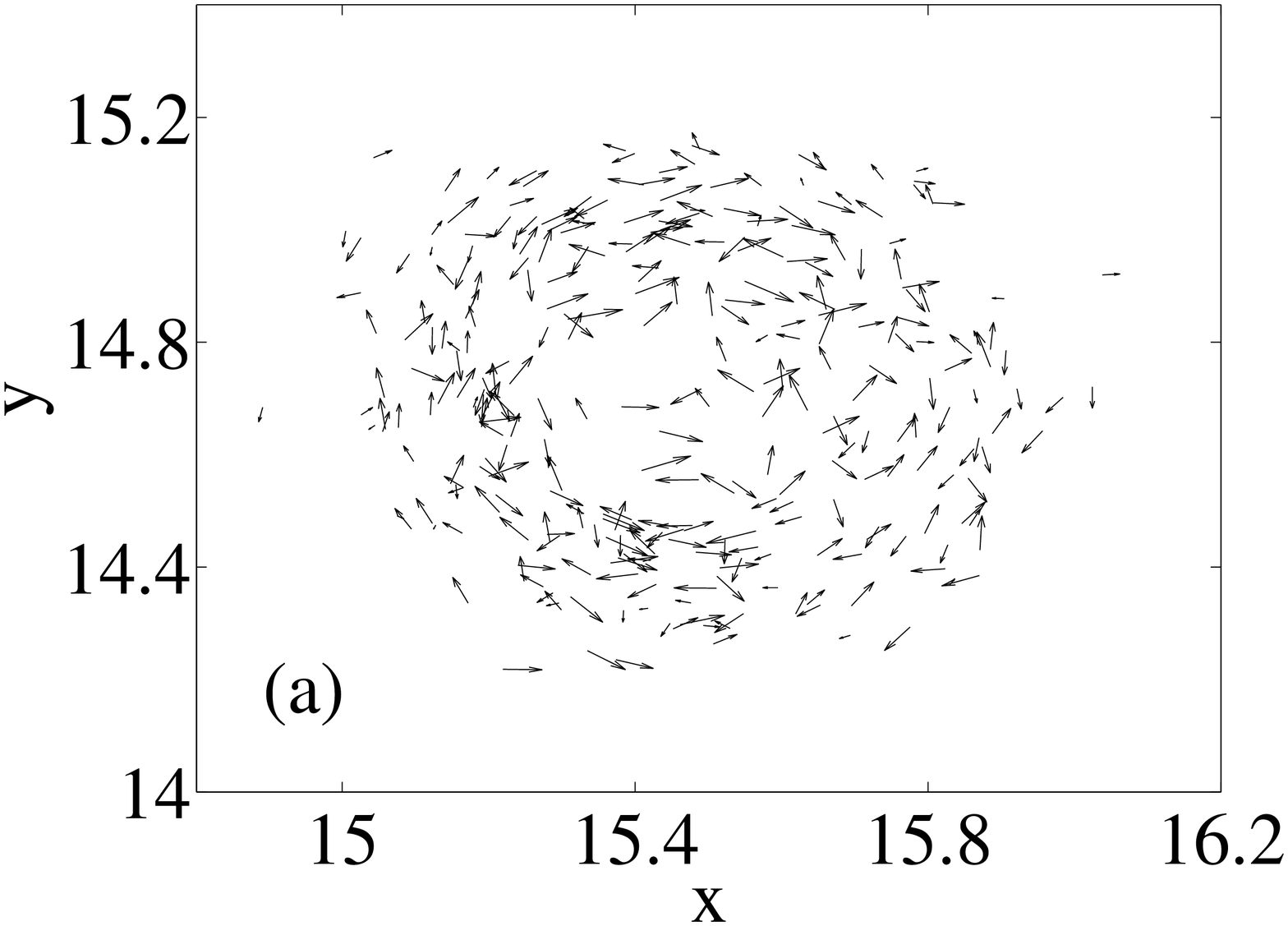}
\end{minipage}
\begin{minipage}{0.49\linewidth}
\includegraphics[width=4.3cm,height=3.0cm]{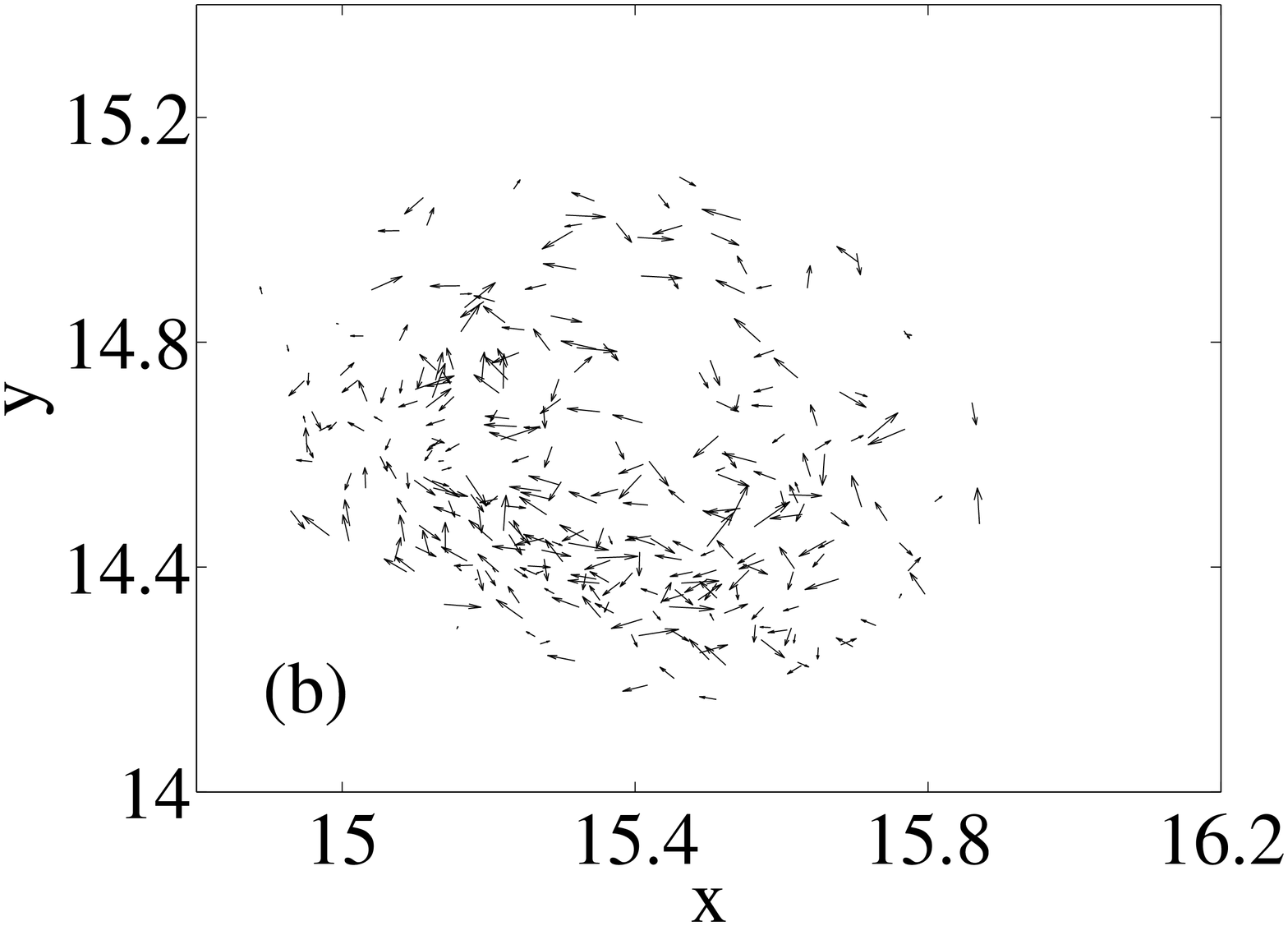}
\end{minipage}\\
\begin{minipage}{0.49\linewidth}
\includegraphics[width=4.3cm,height=3.0cm]{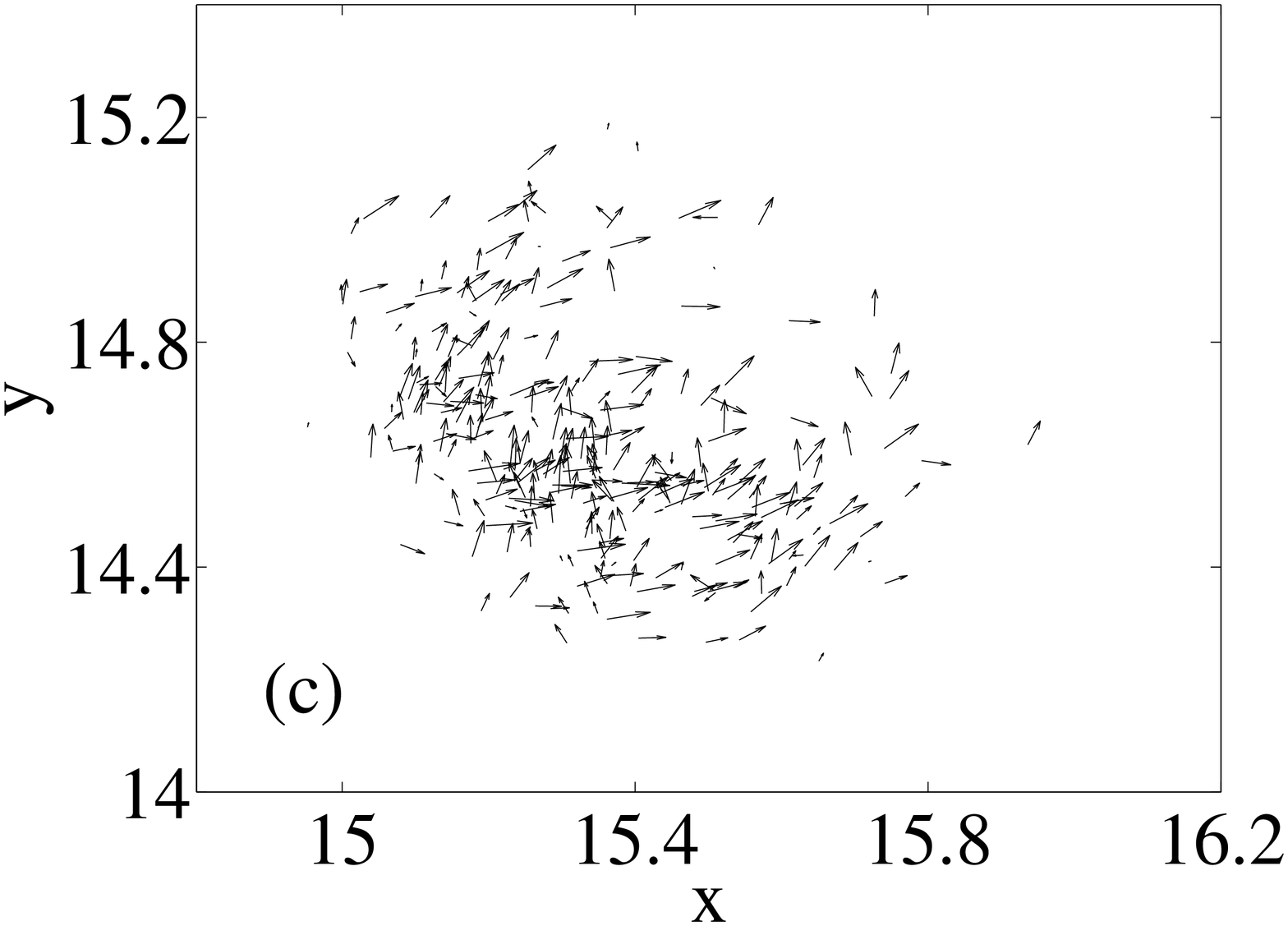}
\end{minipage}
\begin{minipage}{0.49\linewidth}
\includegraphics[width=4.3cm,height=3.0cm]{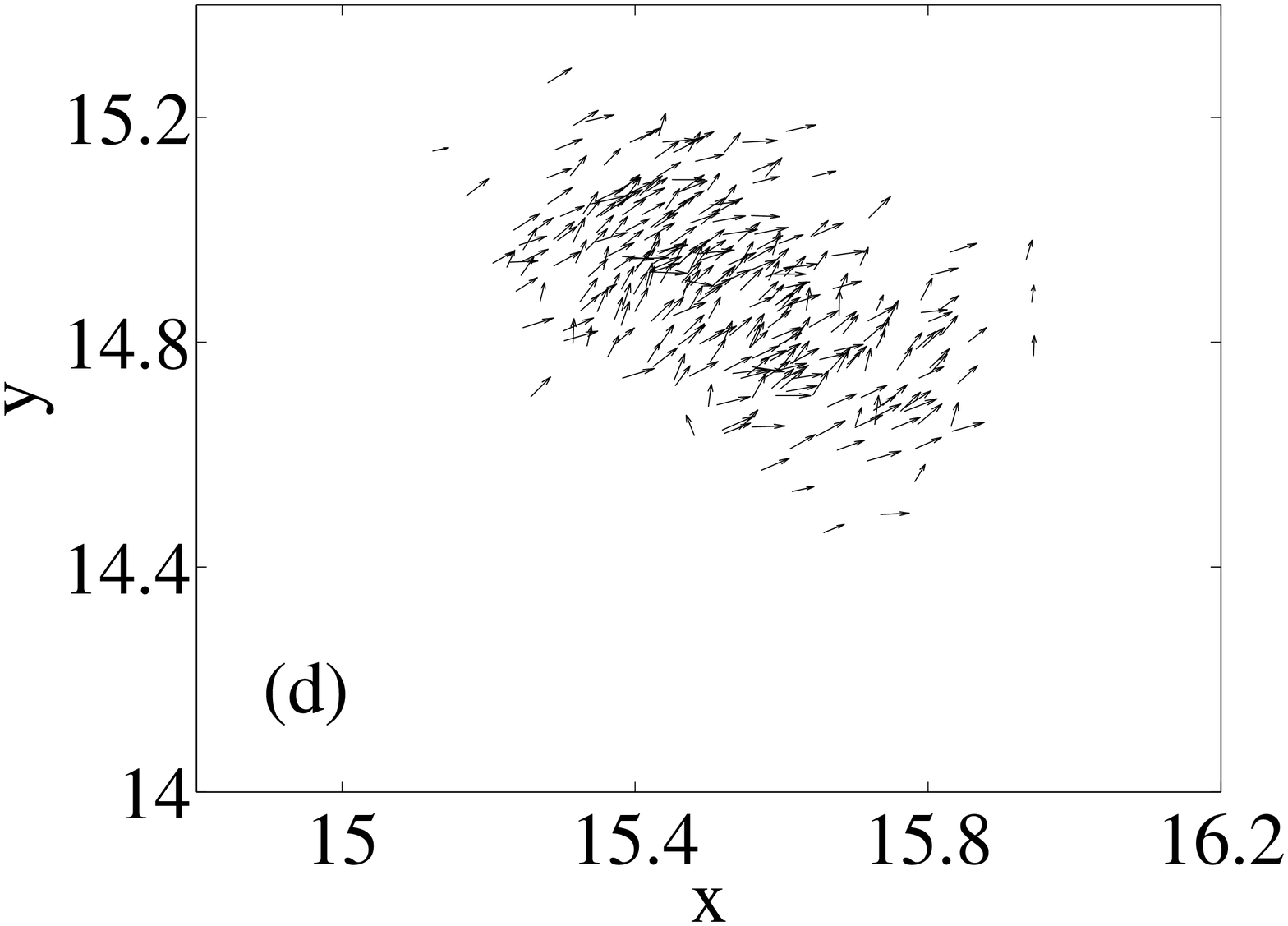}
\end{minipage}\\
\begin{minipage}{0.49\linewidth}
\includegraphics[width=4.3cm,height=3.0cm]{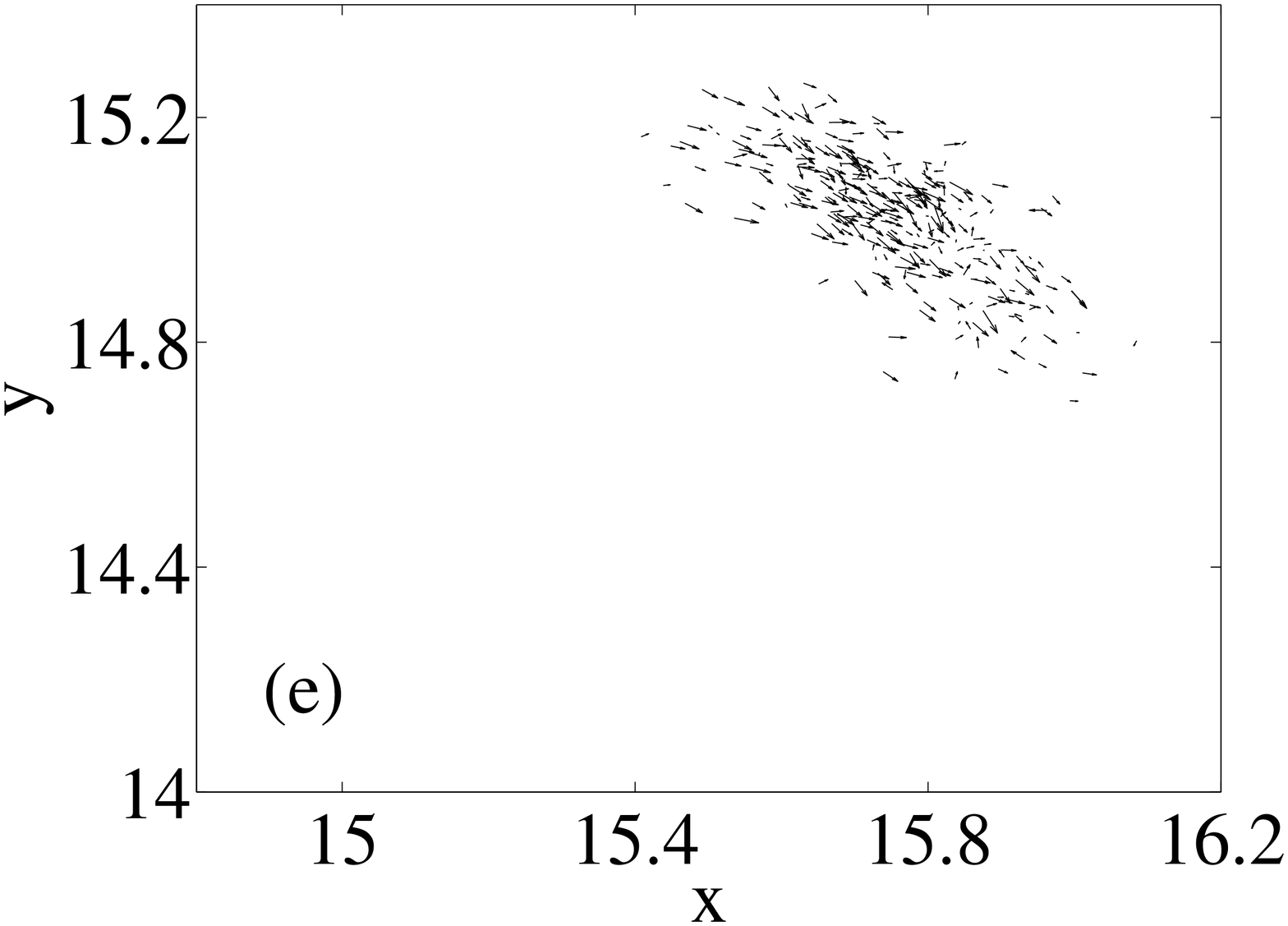}
\end{minipage}
\begin{minipage}{0.49\linewidth}
\includegraphics[width=4.3cm,height=3.0cm]{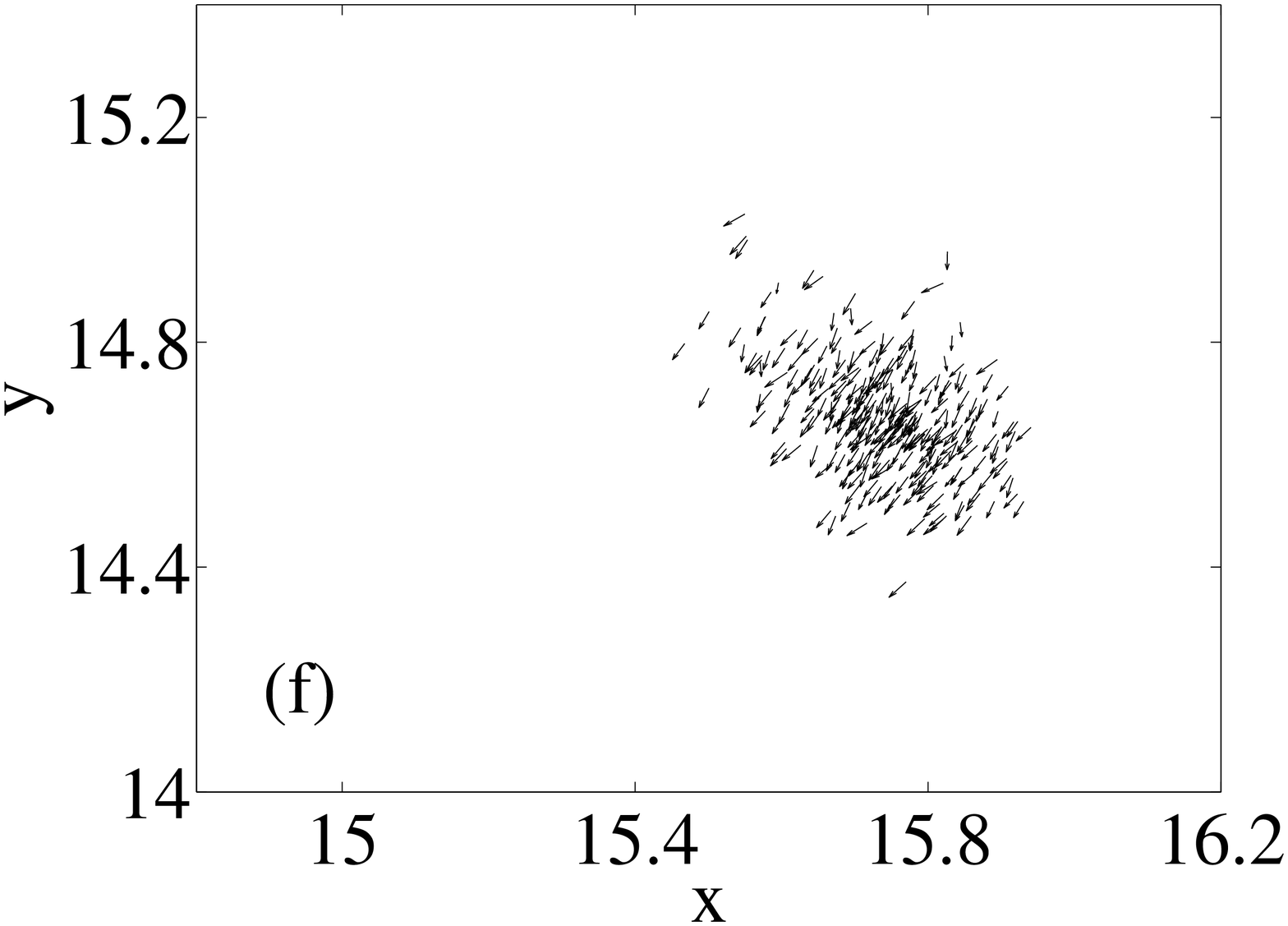}
\end{minipage}\\
\begin{minipage}{0.49\linewidth}
\includegraphics[width=4.3cm,height=3.0cm]{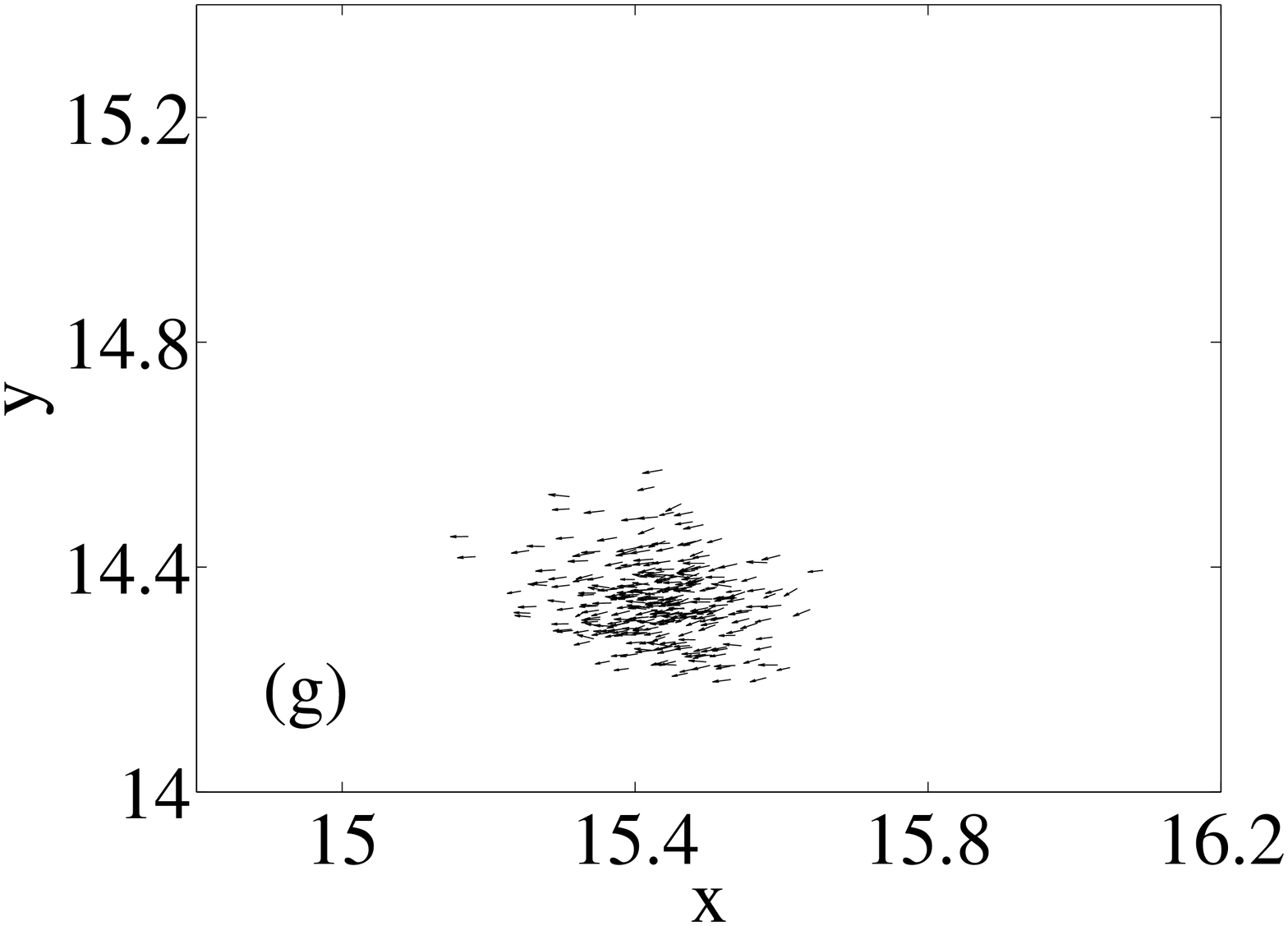}
\end{minipage}
\begin{minipage}{0.49\linewidth}
\includegraphics[width=4.3cm,height=3.0cm]{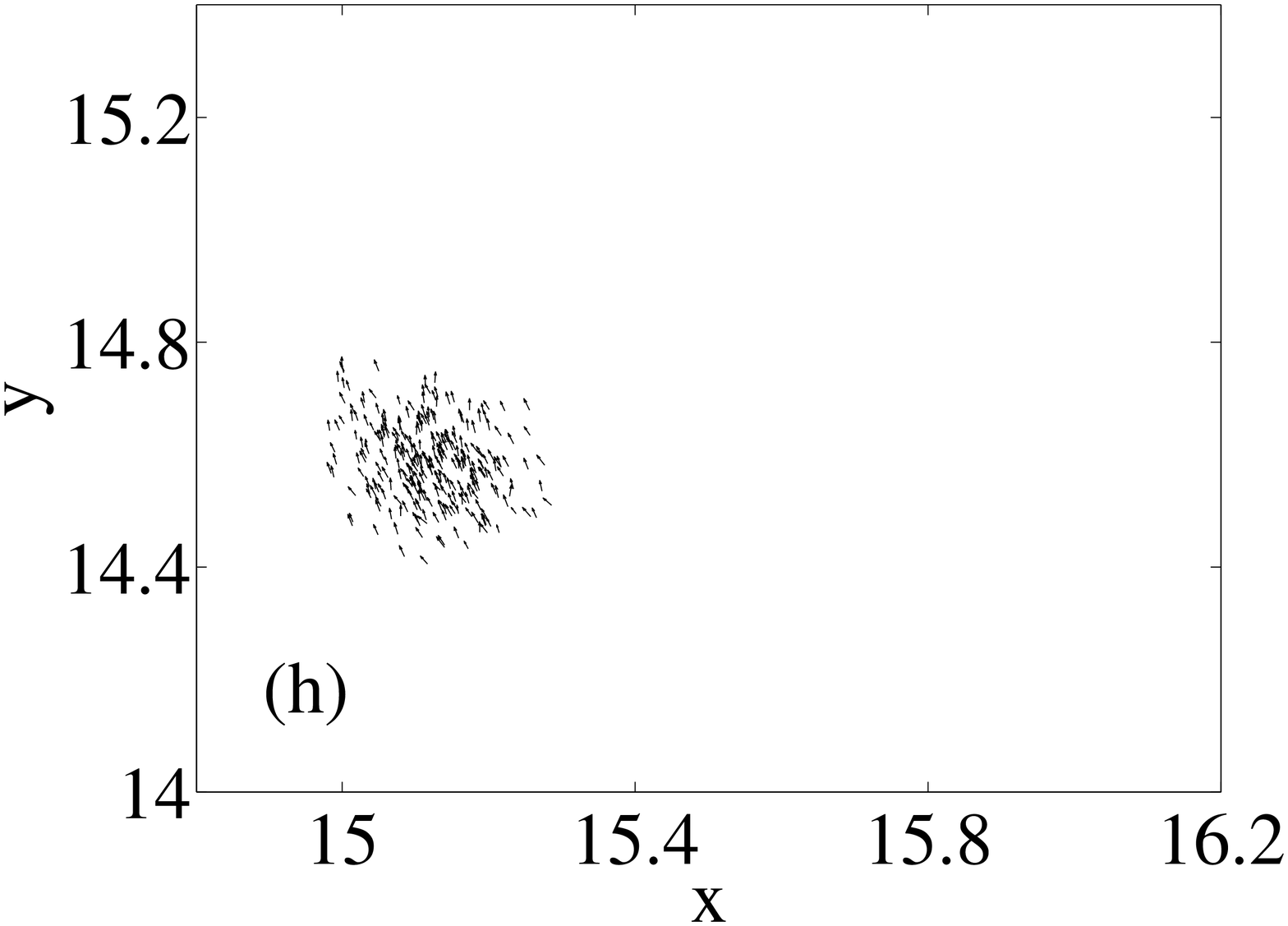}
\end{minipage}\\
\begin{minipage}{0.49\linewidth}
\includegraphics[width=4.3cm,height=3.0cm]{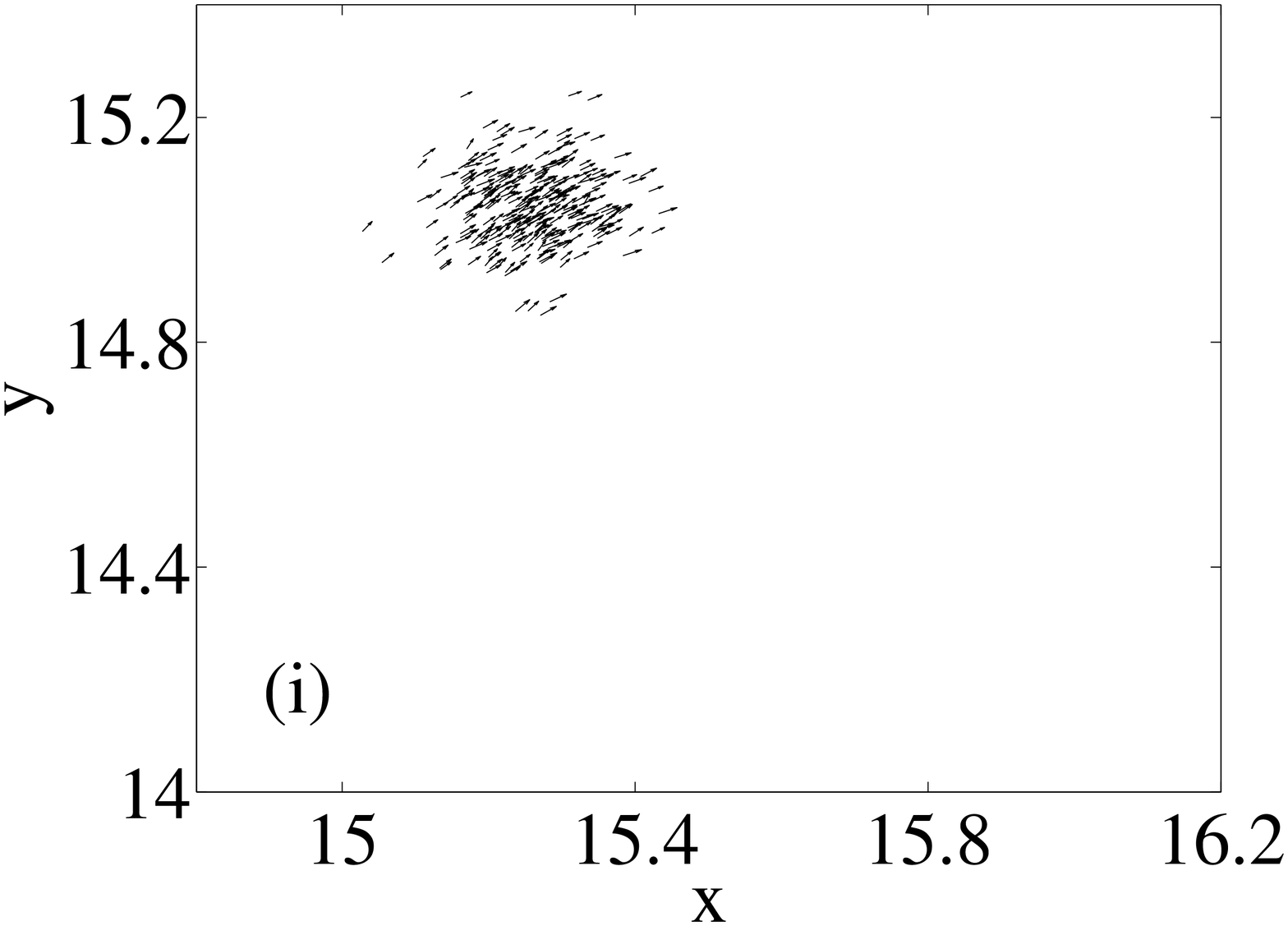}
\end{minipage}
\begin{minipage}{0.49\linewidth}
\includegraphics[width=4.3cm,height=3.0cm]{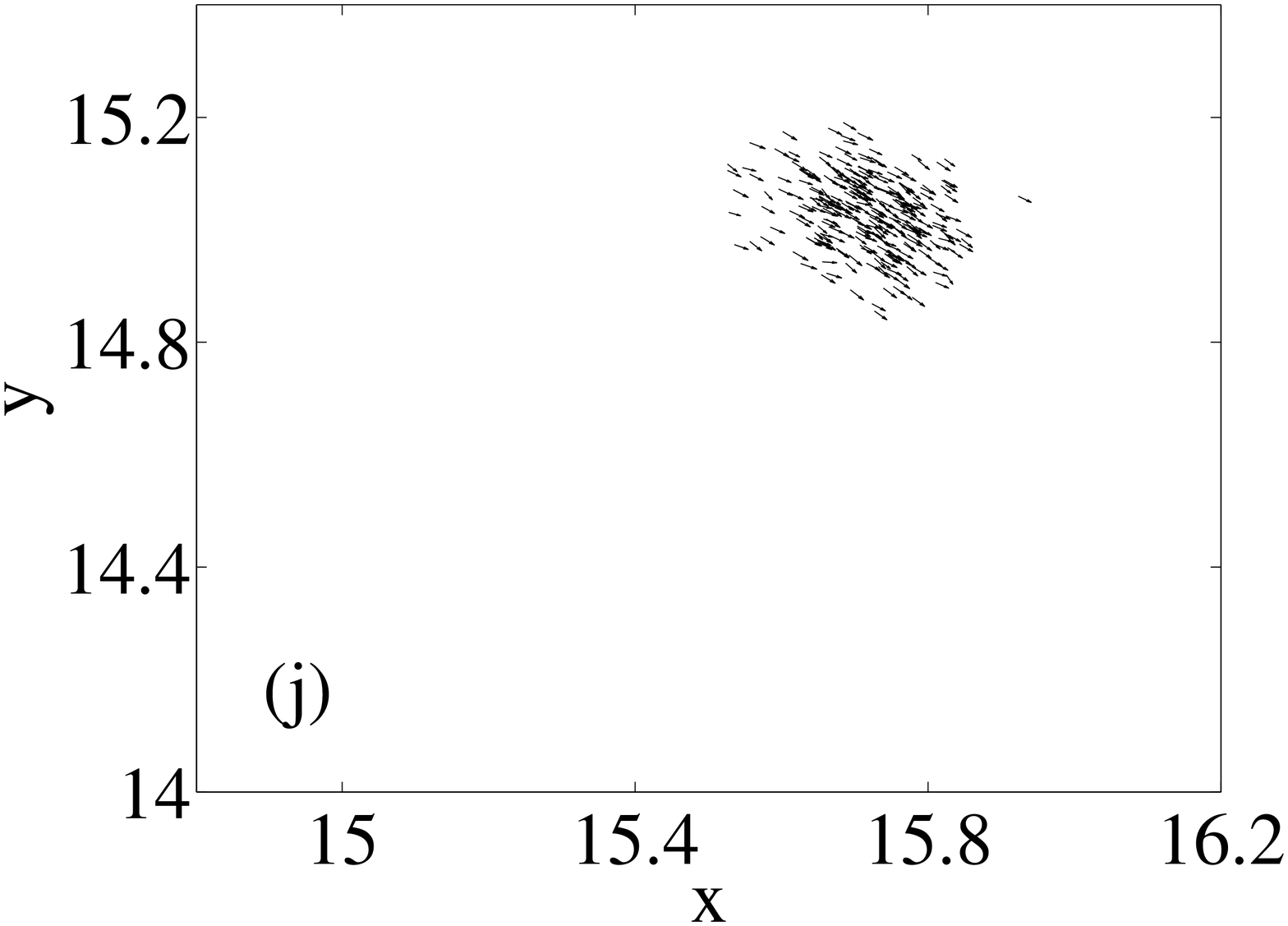}
\end{minipage}
\caption{\label{fig:delay_a4} Snapshots of a swarm taken at (a) $t=50$, (b)
  $t=60$, (c) $t=62$, (d) $t=64$, (e) $t=66$, (f) $t=68$, (g) $t=70$, (h)
  $t=72$, (i) $t=74$, and (j) $t=76$, with $a=4$, $N=300$, and $D=0.08$.  The
  swarm was in a rotational state when the time delay of $\tau=1$ was switched
  on at $t=40$. For a movie, see the relevant mpeg video. Figure reproduced with permission from \cite{Forgoston08}.}
\end{figure}

\section{The Swarm Model}
We consider a two-dimensional swarm with $N$ self-propelling particles that
are mutually attracted in a symmetric fashion. { Additionally,} we consider the case
in which particles communicate with each other with a time delay. The
swarm is governed by the following system of  ODEs:
\begin{subequations}
\begin{align}
\dot{\mathbf{r}}_i =& \mathbf{v}_i,\label{e:swarm_eq_a}\\
\dot{\mathbf{v}}_i =& \left(1 - |\mathbf{v}_i|^2\right)\mathbf{v}_i -
\frac{a}{N}\mathop{\sum_{j=1}^N}_{i\neq j}(\mathbf{r}_i(t) -
\mathbf{r}_j(t-\tau)) + \boldsymbol{\eta}_i(t),\label{e:swarm_eq_b}
\end{align}
\end{subequations}
for $i =1,2\ldots,N$. Here $\mathbf{r}_i$ and $\mathbf{v}_i$  represent the
position and velocity of the
$i$-th particle, respectively; the strength of the attraction is measured by the coupling
constant $a$ and the time delay is uniform and  given by $\tau$. The self-{
propulsion and frictional drag} 
on each particle is given by the term $\left(1 -
  |\mathbf{v}_i|^2\right)\mathbf{v}_i$.  In the absence of coupling, particles
tend to move on a straight line with unit speed $|\mathbf{v}_i| = 1$ as time
goes to infinity. The term $\boldsymbol{\eta}_i(t) = (\eta_i^{(1)}, \eta_i^{(2)})$ is a two-dimensional
vector of stochastic white noise with intensity equal to $D$ and correlation
functions $\langle \eta_i^{(\ell)}(t)\rangle=0$ and $\langle \eta_i^{(\ell)}(t)
\eta_j^{(k)}(t') \rangle = 2D\delta(t-t')\delta_{ij}\delta_{\ell k}$ for
$i,j=1,2,\ldots N$ and $\ell, k = 1,2$.

The coupling between particles arises from a time-delayed, spring-like
potential. Hence, our equations of motion may be considered to be the first
term in a Taylor expansion of other more general time-delayed potential
functions about an equilibrium point.   { The model described by
  Eqs.~(\ref{e:swarm_eq_a})-(\ref{e:swarm_eq_b}) with $\tau=0$ (i.e. no time
  delay) possesses a noise-induced transition whereby a large enough noise
  intensity causes a translating swarm of individuals to transition to a
  rotating swarm with a stationary center of mass~\cite{eem05,Forgoston08}. } Regardless of which state the swarm is in (translating or rotating), the
addition of a communication time delay leads to another type of transition.
This transition occurs if the coupling parameter $a$, is large enough.  As an
example, we consider a swarm that has already undergone a noise-induced
transition to a rotational state before switching on the communication time
delay.  

Figures~\ref{fig:delay_a4}(a)-\ref{fig:delay_a4}(j) show
snapshots of a swarm at $t=50$, $t=60$, $t=62$, $t=64$, $t=66$, $t=68$,
$t=70$, $t=72$, $t=74$, and $t=76$ respectively. For these simulations,
$N=300$, $\tau=1$, $D=0.08$, the noise was switched on at $t=10$ (causing the
swarm to transition to a stationary, rotating state), and once in this
rotating state, the time delay was switched on at $t=40$. One can see that
with the evolution of time, the individual particles become aligned with one
another and the swarm becomes more compact.  Additionally, the swarm is no
longer stationary, but has begun to oscillate
[Figs.~\ref{fig:delay_a4}(g)-\ref{fig:delay_a4}(j)].  

This compact, oscillating aligned swarm state looks similar to
a single ``clump'' that is described in~\cite{dcbc06}.  However, where each
``clump'' of~\cite{dcbc06} contains only some of the total number of swarming
particles, our swarm contains every particle.  Additionally, while a
deterministic model along with global coupling is used to attain the
``clumps'' of~\cite{dcbc06}, our oscillating aligned swarm is attained with the use of
noise and a time delay. 
\section{Mean Field Approximation}

{ As we have shown, once the stochastic swarm is in the stationary,
  rotating state, the addition of a time delay induces an instability.  We
   investigate the stability of the swarm by deriving the mean field
  equations and performing a bifurcation analysis.} 

We carry out a mean field approximation of the swarming system by switching to
particle coordinates relative to the center of mass and disregarding the noise
terms. The center of mass of the swarming system is given by
\begin{align}
\mathbf{R}(t) = \frac{1}{N} \sum_{i=1}^N\mathbf{r}_i(t).
\end{align}
We decompose the position of each particle into
\begin{align}\label{pos_decomp}
\mathbf{r}_i = \mathbf{R} + \delta \mathbf{r}_i,  \qquad i =1,2\ldots,N,
\end{align}
where 
\begin{align}\label{linear_dep}
\sum_{i=1}^N\delta\mathbf{r}_i(t) = 0.
\end{align}
 Inserting Eq.~\eqref{pos_decomp} into the second order system
 equivalent to Eqs.~(\ref{e:swarm_eq_a})-(\ref{e:swarm_eq_b}) with $D=0$ and
 simplifying  { one obtains}
\begin{align}\label{CM1}
\ddot{\mathbf{R}} + \delta\ddot{\mathbf{r}}_i =& \left(1 - |\dot{\mathbf{R}}|^2 -
2\dot{\mathbf{R}}\cdot \delta\dot{\mathbf{r}}_i -
|\delta\dot{\mathbf{r}}_i|^2\right)(\dot{\mathbf{R}} +
\delta\dot{\mathbf{r}_i})\notag\\
& - \frac{a(N-1)}{N}\bigg(\mathbf{R}(t) - \mathbf{R}(t-\tau) +
\delta\mathbf{r}_i(t)\bigg) \notag\\
&- \frac{a}{N}\delta\mathbf{r}_i(t-\tau),
\end{align}
where we used Eq.~\eqref{linear_dep} in the form 
\begin{equation}
\delta\mathbf{r}_i(t-\tau) =
-\sum_{j=1, \ i\neq j}^N \delta\mathbf{r}_j(t-\tau).
\end{equation}
 Summing Eq.~\eqref{CM1} over $i$ and using Eq.~\eqref{linear_dep}, { one arrives at}
\begin{align}\label{CM}
\ddot{\mathbf{R}}=& \left(1 - |\dot{\mathbf{R}}|^2
  -\frac{1}{N}\sum_{i=1}^N|\delta\dot{\mathbf{r}}_i|^2\right)\dot{\mathbf{R}}
\notag\\
&- \frac{1}{N}\sum_{i=1}^N\left(2\dot{\mathbf{R}}\cdot \delta\dot{\mathbf{r}}_i +
|\delta\dot{\mathbf{r}}_i|^2\right)\delta\dot{\mathbf{r}_i}   \notag\\
& -a\frac{N-1}{N}\left(\mathbf{R}(t) - \mathbf{R}(t-\tau)\right).
\end{align}

 { Inserting} Eq.~\eqref{CM} into Eq.~\eqref{CM1} {the following} equation for
$\delta \ddot{\mathbf{r}}_i$ { is obtained}:
\begin{align}\label{dri}
\delta\ddot{\mathbf{r}}_i=&
\left(\frac{1}{N}\sum_{j=1}^N|\delta\dot{\mathbf{r}}_j|^2 -
2\dot{\mathbf{R}}\cdot \delta\dot{\mathbf{r}}_i -  |\delta\dot{\mathbf{r}}_i|^2
\right)\dot{\mathbf{R}} \notag\\
&+ \left(1 - |\dot{\mathbf{R}}|^2 -
2\dot{\mathbf{R}}\cdot \delta\dot{\mathbf{r}}_i -
|\delta\dot{\mathbf{r}}_i|^2\right)\delta\dot{\mathbf{r}}_i\notag\\
&+\frac{1}{N}\sum_{j=1}^N\left(2\dot{\mathbf{R}}\cdot
\delta\dot{\mathbf{r}}_j + |\delta\dot{\mathbf{r}}_j|^2\right)
\ \delta\dot{\mathbf{r}}_j - a \frac{N-1}{N} \delta\mathbf{r}_i \notag\\
&- \frac{a}{N}\delta\mathbf{r}_i(t-\tau),
\end{align}
for $i =1,2\ldots,N$. 

Equations~\eqref{CM} and~\eqref{dri} are fully equivalent to
Eqs.~(\ref{e:swarm_eq_a})-(\ref{e:swarm_eq_b}) when $D=0$, and simply consist of rewriting the original system
{ using the relationship between the}  particle coordinates
$\mathbf{r}_i$, { the} center of mass $\mathbf{R}$, and {
  the coordinates}
relative to the center of mass  $\delta\mathbf{r}_i$. This mapping has
transformed the original $2N$ differential equations into $2N+2$ {
  differential equations}. There is,
however, no inconsistency since in the transformed set of equations only $2N$
of them are independent, because of the relation { seen} in Eq.~\eqref{linear_dep}.

We then obtain a mean field approximation by neglecting the fluctuation of the
swarm particles, {$\delta\mathbf{r}_i$'s}, from the center of mass:
\begin{align}\label{mean_field}
\ddot{\mathbf{R}}=& \left(1 - |\dot{\mathbf{R}}|^2 \right)\dot{\mathbf{R}} -a\left(\mathbf{R}(t) - \mathbf{R}(t-\tau)\right),
\end{align}
where we made the approximation $a\frac{N-1}{N}\approx a$ since we consider the thermodynamic limit. 

\section{Bifurcations in the Mean Field Equation}

The behavior of the system in the mean field approximation in different
regions of parameter space may be better understood by using bifurcation
analysis. Equation~\eqref{mean_field} may be written in component form using 
$\mathbf{R} = (X, Y)$ and $\dot{\mathbf{R}} = (U, V)$ as 
\begin{subequations}
\begin{align}
\dot{X} &= U,\label{e:CM_components_a}\\
\dot{U} &= (1 - U^2 - V^2)U - a(X - X(t -\tau)),\\
\dot{Y} &= V,\\
\dot{V} &= (1 - U^2 - V^2)V - a(Y - Y(t-\tau)).\label{e:CM_components_d}
\end{align}
\end{subequations}
For all values of $a$ and $\tau$, Eqs.~(\ref{e:CM_components_a})-(\ref{e:CM_components_d}) have
translationally invariant stationary solutions
\begin{align}
X = X_0, \quad U = 0, \quad Y = Y_0, \quad V=0,
\end{align}
with two free parameters $X_0$ and $Y_0$. They also have a three parameter family of uniformly translating solutions
\begin{align}
X = U_0 t + X_0, \quad U = U_0, \quad Y = V_0 t + Y_0, \quad V = V_0,
\end{align}
which requires
\begin{align}
 U_0^2 + V_0^2 = 1 - a\tau,
\end{align}
and thus exists only for $a\tau < 1$. In the two-parameter space $(a, \tau)$,
the hyperbola $a \tau = 1$ is in fact a pitchfork bifurcation line on which
the uniformly translating states are born from the stationary state $(X_0, 0, Y_0,
0)$. The other branch of the pitchfork is an unphysical solution with
negative speed.

Linearizing Eqs.~(\ref{e:CM_components_a})-(\ref{e:CM_components_d}) about the stationary state, we obtain the
characteristic equation
\begin{align}
\left( a(1- e^{-\lambda\tau}) - \lambda + \lambda^2 \right)^2 = 0.
\end{align}
 It suffices to study the zeros of the function 
\begin{align}\label{char_eq}
{\cal{D}}(\lambda) = a(1- e^{-\lambda\tau}) - \lambda + \lambda^2 = 0,
\end{align}
since the eigenvalues of
  Eqs.~(\ref{e:CM_components_a})-(\ref{e:CM_components_d}) are obtained by
  duplicating those of Eq.~\eqref{char_eq}.

We now search for Hopf bifurcations in the two parameter space $(a,
\tau)$ by { letting} $\lambda = i \omega$ in
Eq.~\eqref{char_eq}. { Substitution leads to}
\begin{align}\label{hopf_cond}
a - \omega^2 - i\omega = a e^{-i\omega \tau}.
\end{align}
Taking the modulus of Eq.~\eqref{hopf_cond}, we  { find} $a$ at
the Hopf point, $a_H$, { is given by}
\begin{align}
a_H^2 = (a_H - \omega^2)^2 + \omega^2,
\end{align}
or, considering $\omega \neq 0$,
\begin{align}\label{a_H}
a_H = \frac{1 + \omega^2}{2}.
\end{align}

\begin{figure}[t!]
\begin{center}
\subfigure{\includegraphics[scale=0.42]{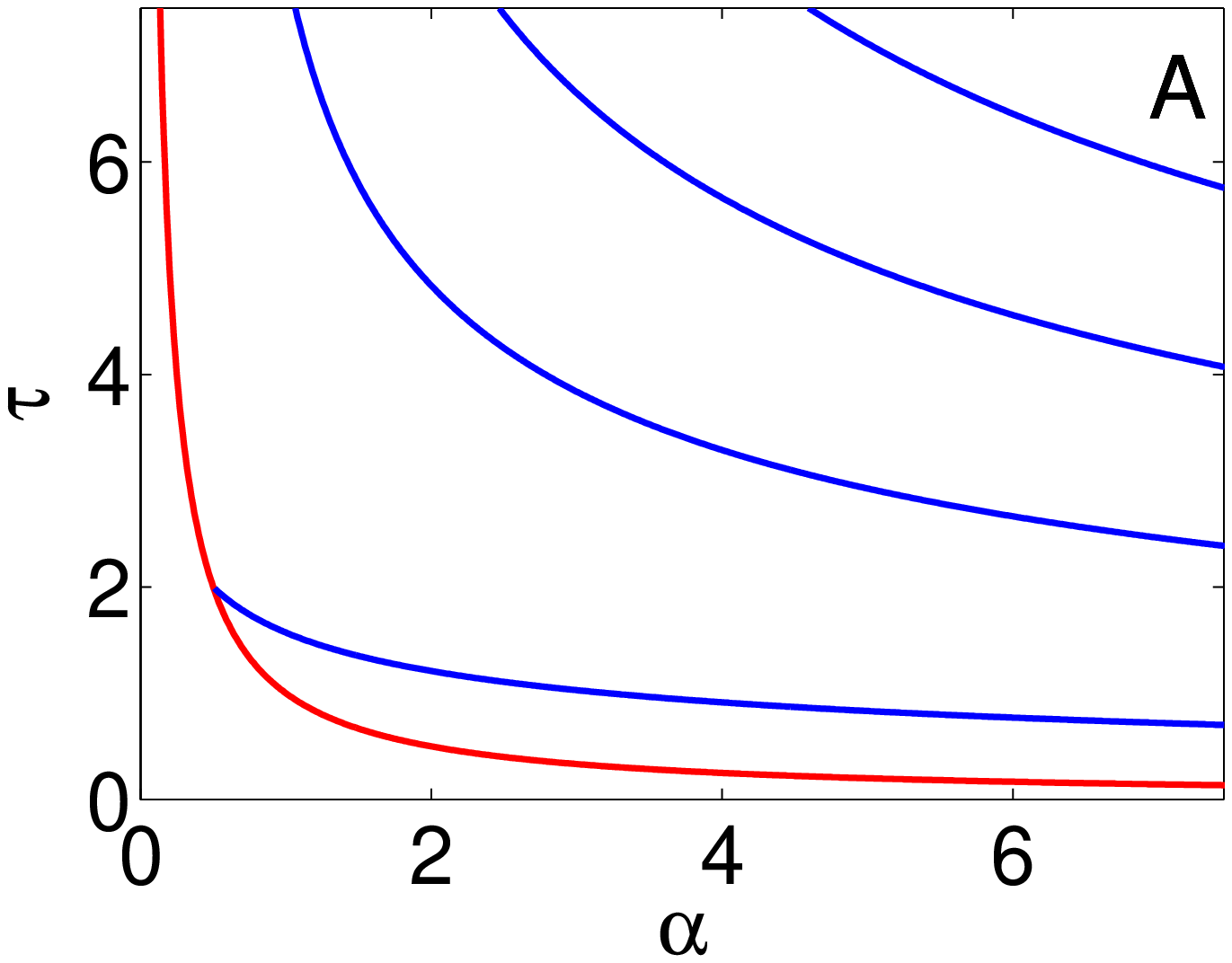} \label{Hopf_pitchfork_alpha_tau}}
\subfigure{\includegraphics[scale=0.42]{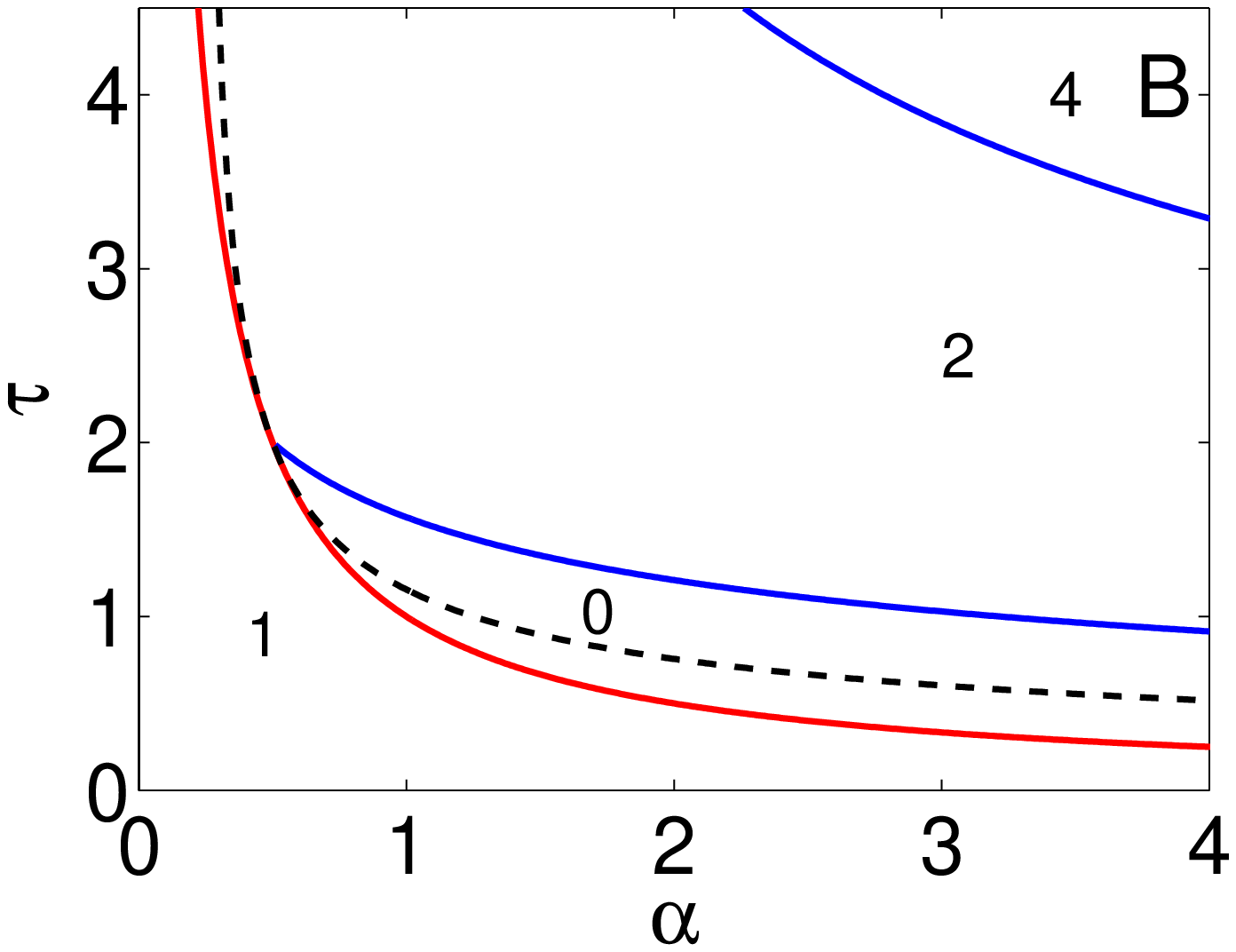} \label{Eigvls_alpha_tau}}
\caption{(a) Hopf (blue) and pitchfork (red) bifurcation curves in $a$ and
  $\tau$ space. (b) A zoom-in of the branches in the first panel displaying also the saddle to node transition (dashed black); the
number in each region indicates the number of eigenvalues with a real part
greater than zero with the solid lines as boundaries. (Color online.)}
\end{center}
\end{figure}

We eliminate $a$ in Eq.~\eqref{hopf_cond} by using Eq.~\eqref{a_H} and
 { taking} the complex conjugate to obtain an equation for $\tau$ at the Hopf point
\begin{align}
\frac{1 - \omega^2}{1+\omega^2} + i\frac{2\omega}{1 + \omega^2} =  e^{i\omega \tau}.\label{e:above}
\end{align}
We obtain $\tau$ by equating the arguments of both sides, being careful to
use the branch of $\tan\theta$ in $(0,\pi)$ since the left hand side of  { Eq.~(\ref{e:above})} is on the upper complex plane for $\omega > 0$.  {The result is} a family of Hopf bifurcation curves parameterized by $\omega$:
\begin{subequations}
\begin{align}
a_H(\omega) &= \frac{1 + \omega^2}{2},\label{e:hopf_omega_a}\\
\tau_{Hn}(\omega) &=
\frac{1}{\omega}\left(\arctan\left(\frac{2\omega}{1-\omega^2}\right) +
2n\pi\right)  \notag \label{e:hopf_omega_b}\\
n &= 0, 1,\ldots.
\end{align}
\end{subequations}
These curves are shown in Fig.~\ref{Hopf_pitchfork_alpha_tau}. We may eliminate the parameter $\omega$ between these two equations and obtain

\begin{align}\label{hopf_a}
\tau_{Hn}(a) &=
\frac{1}{\sqrt{2a -1}}\left(\arctan\left(\frac{\sqrt{2a-1}}{1-a}\right) +
2n\pi\right) \notag\\ 
n &= 0, 1,\ldots
\end{align}
In spite of their appearance, the Hopf curves in Eqs.~(\ref{e:hopf_omega_a})-(\ref{e:hopf_omega_b}) and~\eqref{hopf_a} are in fact continuous at $\omega = 1$ and $a=1$,
respectively (with the correct branch of $\tan\theta$ in $(0, \pi)$).
 From Eq.~(\ref{e:hopf_omega_a})-(\ref{e:hopf_omega_b}), we see that the Hopf
frequency depends only on the value of $a$ for all members in the family; it
has the value one at $a=1$ and the frequency tends to infinity as
$a$ grows. Interestingly, only the first Hopf curve of the family in
Eq.~\eqref{hopf_a} is defined at $a=1/2$; it has the value $\tau_{H0}\vert_{a=1/2} = 2$. The point ($a=1/2$, $\tau = 2$) which lies both on the first member of
the family of Hopf curves and on the pitchfork branch is in fact a Bogdanov-Takens
(BT) point~\cite{GuckenheimerHolmes83}, where $\omega = 0$. None of the other Hopf branches meet the pitchfork bifurcation line since
they tend asymptotically to infinity as $a\rightarrow 1/2$.

\begin{figure}[t!]
\begin{center}
\subfigure{\includegraphics[scale=0.25]{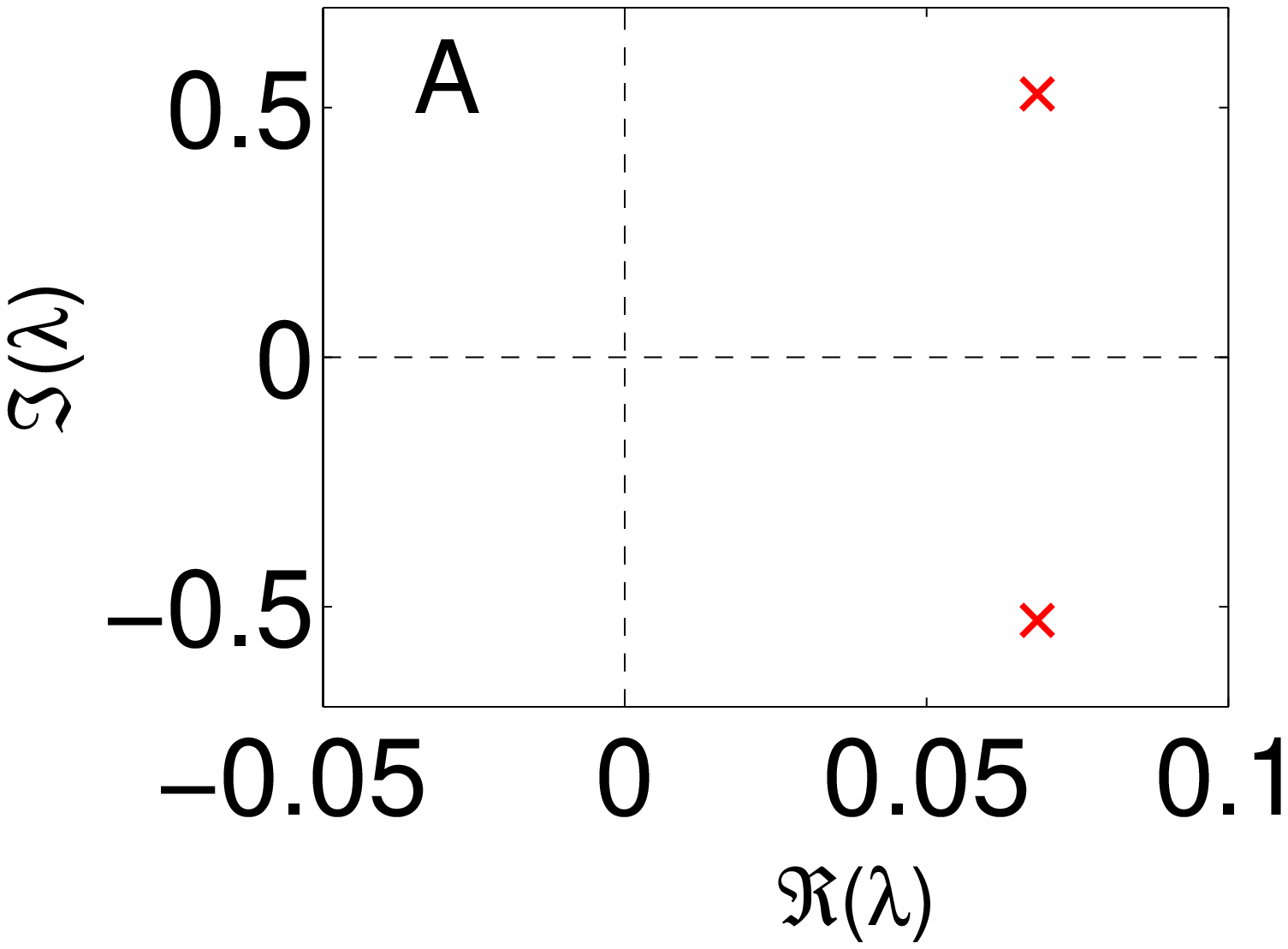} \label{BT_a}}
\subfigure{\includegraphics[scale=0.25]{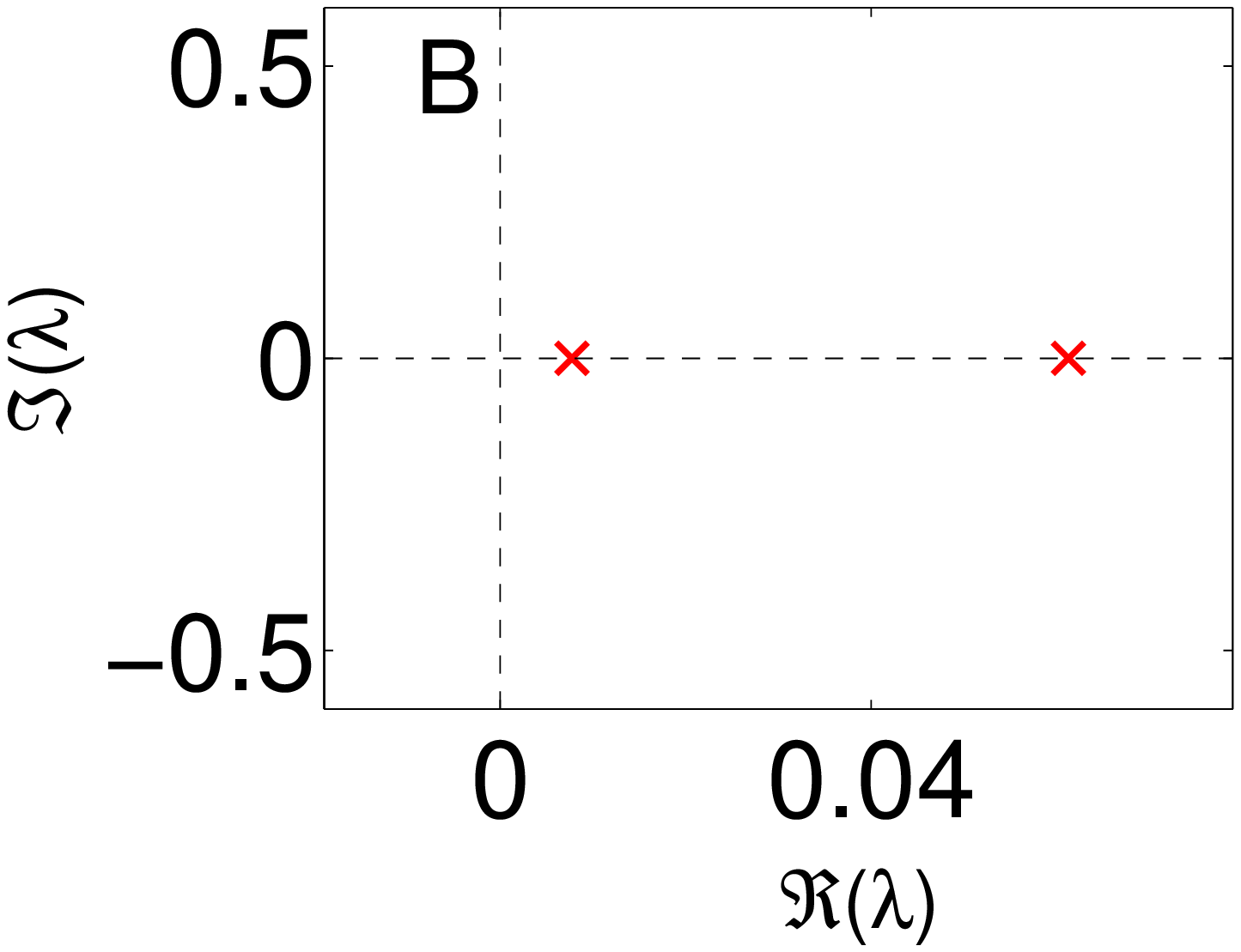} \label{BT_b}}
\subfigure{\includegraphics[scale=0.25]{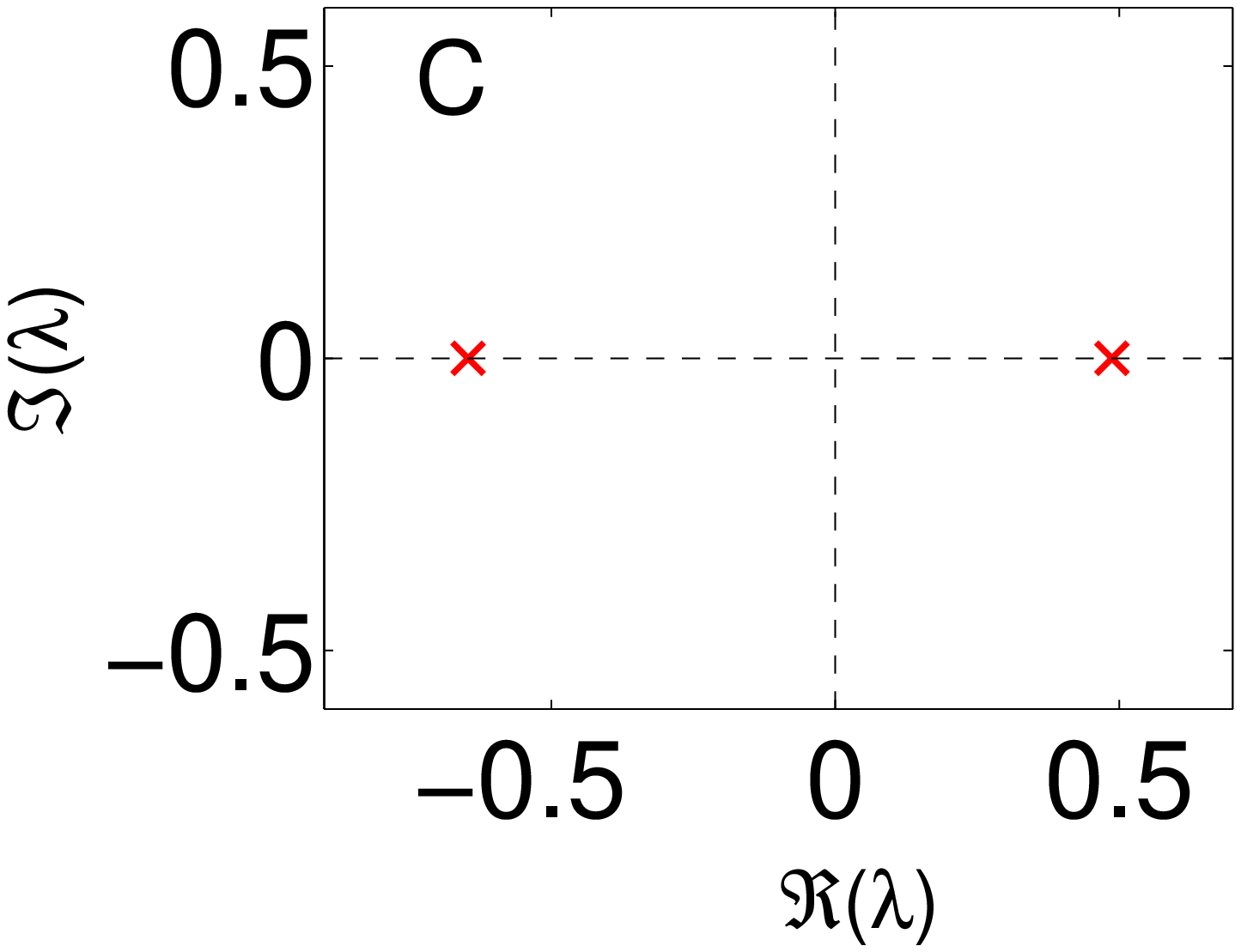} \label{BT_c}}
\subfigure{\includegraphics[scale=0.25]{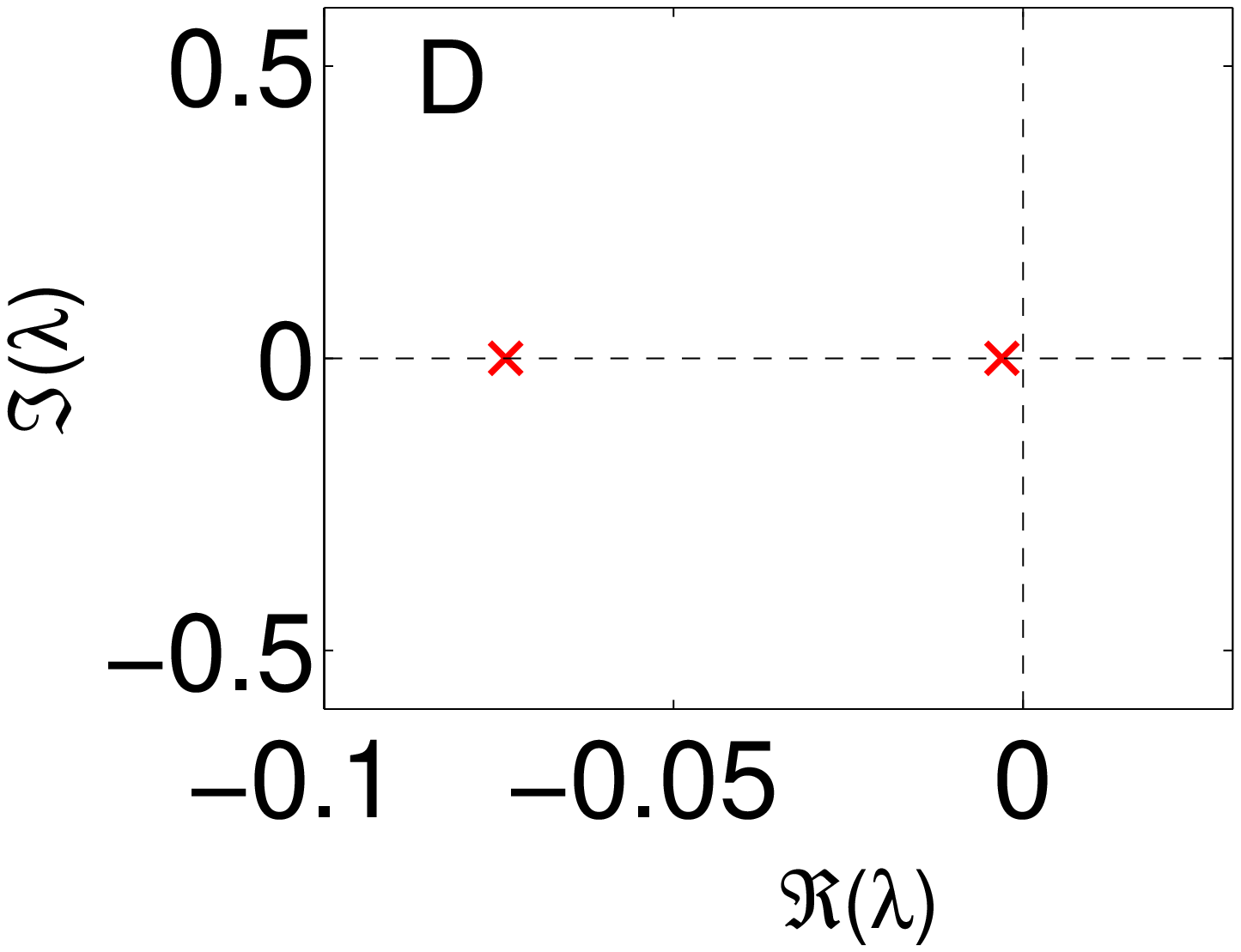} \label{BT_d}}
\subfigure{\includegraphics[scale=0.25]{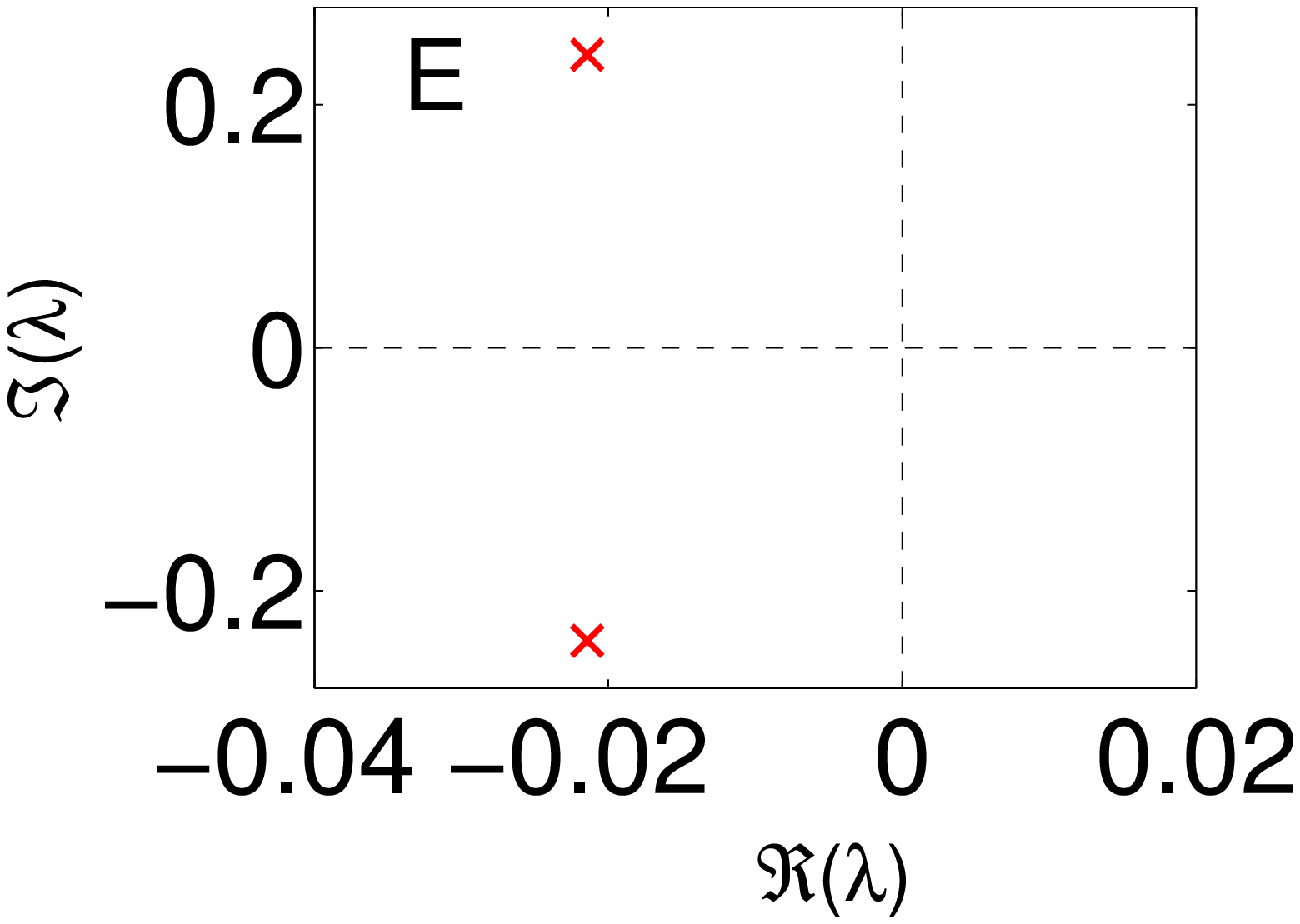} \label{BT_e}}
\caption{Location of the dominating eigenvalues around the Bogdanov-Takens
  point at $a = 1/2$, $\tau = 2$. Parameter values are (a) $a = 0.60$, $\tau
  = 2.0$, (b) $a = 0.48$, $\tau  = 2.09$, (c) $a = 0.40$,
  $\tau  = 2.01$, (d) $a = 0.53$,   $\tau  = 1.90$, and (e) $a = 0.55$, $\tau  = 1.91$.}\label{BT_fig}
\end{center}
\end{figure}

We used a numerical continuation method (DDE-Biftool)~\cite{Engel} to calculate the
pitchfork and Hopf branches in the $(a, \tau)$ parameter space; these
results are in perfect agreement with our analytical calculations (results not
shown). These numerical studies reveal that the number of eigenvalues with real part
greater than zero is as indicated in Fig.~\ref{Eigvls_alpha_tau}. In
addition, our numerical continuation analyses also reveal node to focus transitions of the steady
state. These occur at points where there are two real and equal
eigenvalues, i.e. where ${\cal{D}}(\lambda) = 0$ and ${\cal{D}'}(\lambda) = 0$, for $\lambda$
real. From ${\cal{D}'}(\lambda) = 0$ we obtain $e^{-\tau\lambda} =
\frac{1-2\lambda}{a\tau}$, which we can insert into ${\cal{D}}(\lambda) = 0$ to obtain
\begin{align}\label{node_focus_quadratic}
\lambda^2 - \left(1 - \frac{2}{\tau}\right)\lambda + a - \frac{1}{\tau} = 0,
\end{align}
with solutions $\lambda = \frac{1}{2}\left[1 - \frac{2}{\tau} \pm \sqrt{ 1
    +\frac{4}{\tau^2} -    4a}\right]$. For the roots to be repeated, we
set the discriminant to zero and this gives the curve where the node-focus transitions occur:
\begin{align}\label{node_focus}
\tau = \frac{1}{\sqrt{a - 1/4}}.
\end{align}
Moreover, from the solutions to Eq.~\eqref{node_focus_quadratic} we see that the repeated eigenvalues
have positive real parts if $\tau > 2$ and negative real parts if $\tau <
2$. In Fig.~\ref{Eigvls_alpha_tau}, we
show the pitchfork and Hopf bifurcation curves overlaid with the node-focus
transition curve { given by} Eq.~\eqref{node_focus}.

As seen in Fig.~\ref{Eigvls_alpha_tau}, the pitchfork and Hopf branches,
together with the node-focus transition curves split the area around the BT
point into five different regions. The behavior of the leading
eigenvalues (excluding the one at the origin) as one probes these five regions
 is shown in Figs.~\ref{BT_a}-\ref{BT_e}.  At a point directly to
the right of the BT point in $(a, \tau)$ space, the stationary solution has a pair of
eigenvalues with positive real parts and non-zero imaginary parts
[Fig.~\ref{BT_a}]. Moving counter-clockwise in the $(a, \tau)$ plane, the eigenvalue pair collapses onto
the real line after crossing the upper branch of the node-focus transition
[Fig.~\ref{BT_b}]. Still moving in the same direction in parameter space, we observe two different instances of eigenvalues
crossing the origin: first, the smaller of the two purely real and positive
eigenvalues does so on the upper part of the  pitchfork bifurcation line
[Fig.~\ref{BT_c}] and then the remaining purely real and positive eigenvalue
crosses the origin on the lower part of the bifurcation line
[Fig.~\ref{BT_d}]. Finally, at the node-focus transition line, the two purely
real and negative eigenvalues coincide on the real axis and  acquire
non-zero imaginary parts [Fig.~\ref{BT_e}]. Continuing upwards in parameter
space, the complex pair crosses the imaginary axis on the Hopf bifurcation
curve, giving birth to a stable limit cycle.

\subsection{Numerical simulations}

\begin{figure}[t!]
\begin{center}
\includegraphics[scale=0.3]{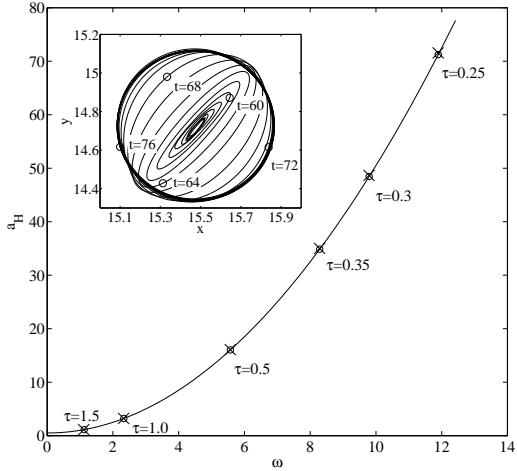} 
\caption{\label{fig:HopfBF} The limit cycle of the center of mass { is
    shown through}  a comparison
  of analytical (solid line) and numerical (``cross'' markers) values of $a_H$
  and $\omega$ for several choices of $\tau$.  The analytical result is found
  using Eqs.~(\ref{e:hopf_omega_a})-(\ref{e:hopf_omega_b}), while the numerical result is found using a
  continuation method~\cite{Engel} for Eq.~(\ref{mean_field}). The inset
    shows the   stochastic trajectory of the center of mass of the swarm from
    $t=45$ to   $t=90$.Figure reproduced with permission from \cite{Forgoston08}.} 
\end{center}
\end{figure}
Figure~\ref{fig:HopfBF} shows an excellent comparison of the analytical result given by
Eqs.~(\ref{e:hopf_omega_a})-(\ref{e:hopf_omega_b}) with a numerical result which was found using
a continuation method~\cite{Engel} for
the mean field model for several choices of
$\tau$.  Furthermore, for $\tau=1$, the value of coupling $a$ at the bifurcation point is
$a_H\approx 3.2$.  This value of $a_H$ corresponds very well to the change in
behavior of the stochastic swarm (results not shown).

More evidence of the Hopf bifurcation is seen in the inset of
Fig.~\ref{fig:HopfBF}.  The inset shows the stochastic trajectory of the center of mass of
the swarm from $t=45$ to $t=90$ for the example shown in Fig.~\ref{fig:delay_a4}.  Once the time delay is switched on at $t=40$
(with the swarm located at the center of the inset figure), the swarm begins
to oscillate.  The swarm moves along an elliptical path [the position of
its center of mass is denoted at several times that correspond to
Figs.~\ref{fig:delay_a4}(b),~\ref{fig:delay_a4}(d),~\ref{fig:delay_a4}(f),~\ref{fig:delay_a4}(h),
and~\ref{fig:delay_a4}(j)], until it eventually converges to the circular
limit cycle. 

\begin{figure}[t!]
\begin{center}
\subfigure{\includegraphics[scale=0.26]{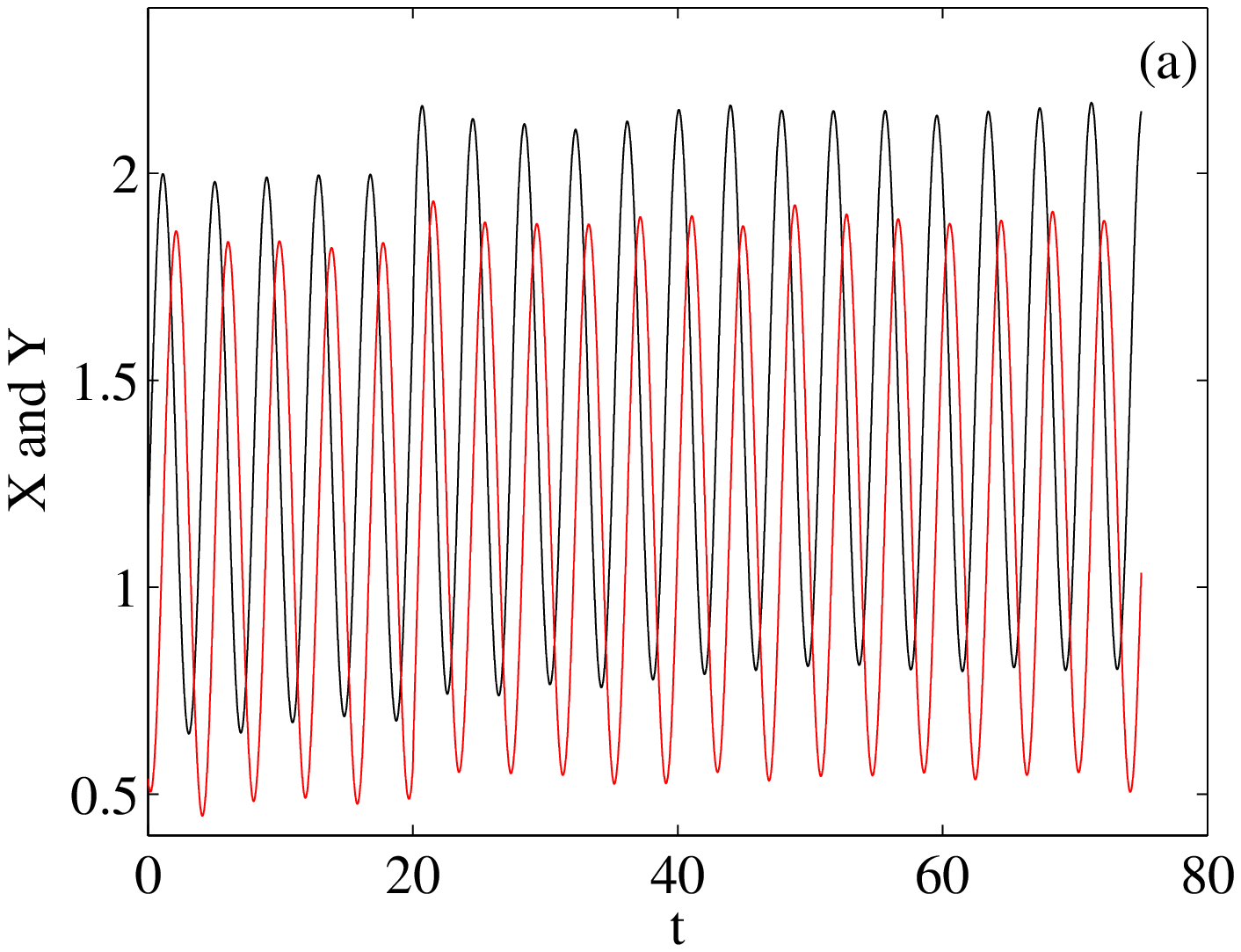} \label{fig:pos_rot}}
\subfigure{\includegraphics[scale=0.26]{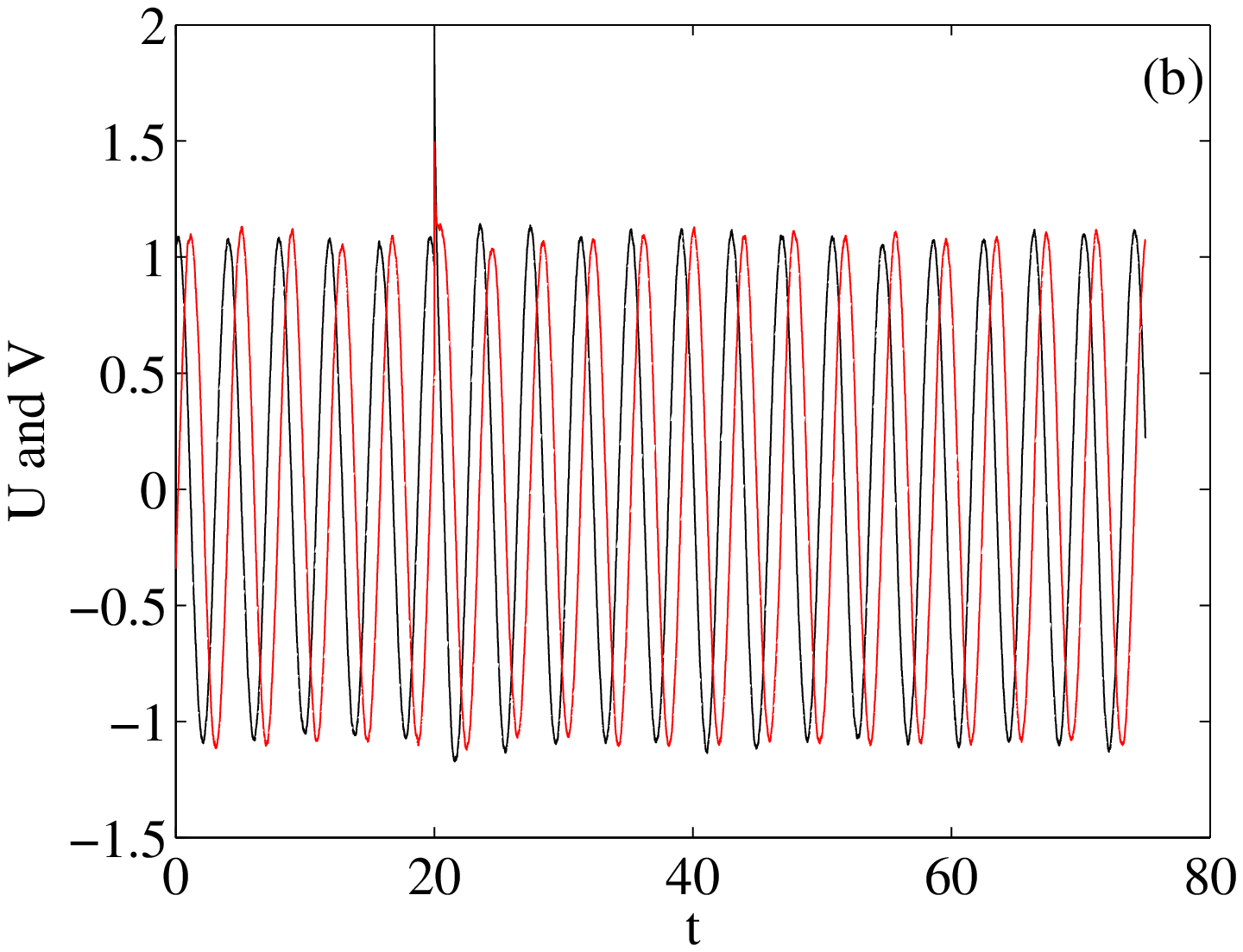} \label{fig:vel_rot}}
\caption{ The limit cycle of the center of mass is
    shown through the (a) position and (b) velocity time
    series of the swarm using $N=75$ particles, $a=0.7$, $\tau =2.2$, and noise intensity $D=0.045$.  A velocity perturbation is applied at $t=20$.}
\end{center}
\end{figure}

{ Figures~\ref{fig:pos_rot} and~\ref{fig:vel_rot} show a}  time series simulation of a
swarm with $N=75$ particles. { Figure~\ref{fig:pos_rot} shows the}
 position  components, {while
  Fig.~\ref{fig:vel_rot} shows the velocity components.},
 { One can see that} the swarm follows a circular-like path
over time. A perturbation { that is applied at $t=20$}
shows that for the chosen parameters, the pattern is stable in the presence of noise.

\begin{figure}[h!]
\begin{center}
\subfigure{\includegraphics[scale=0.26]{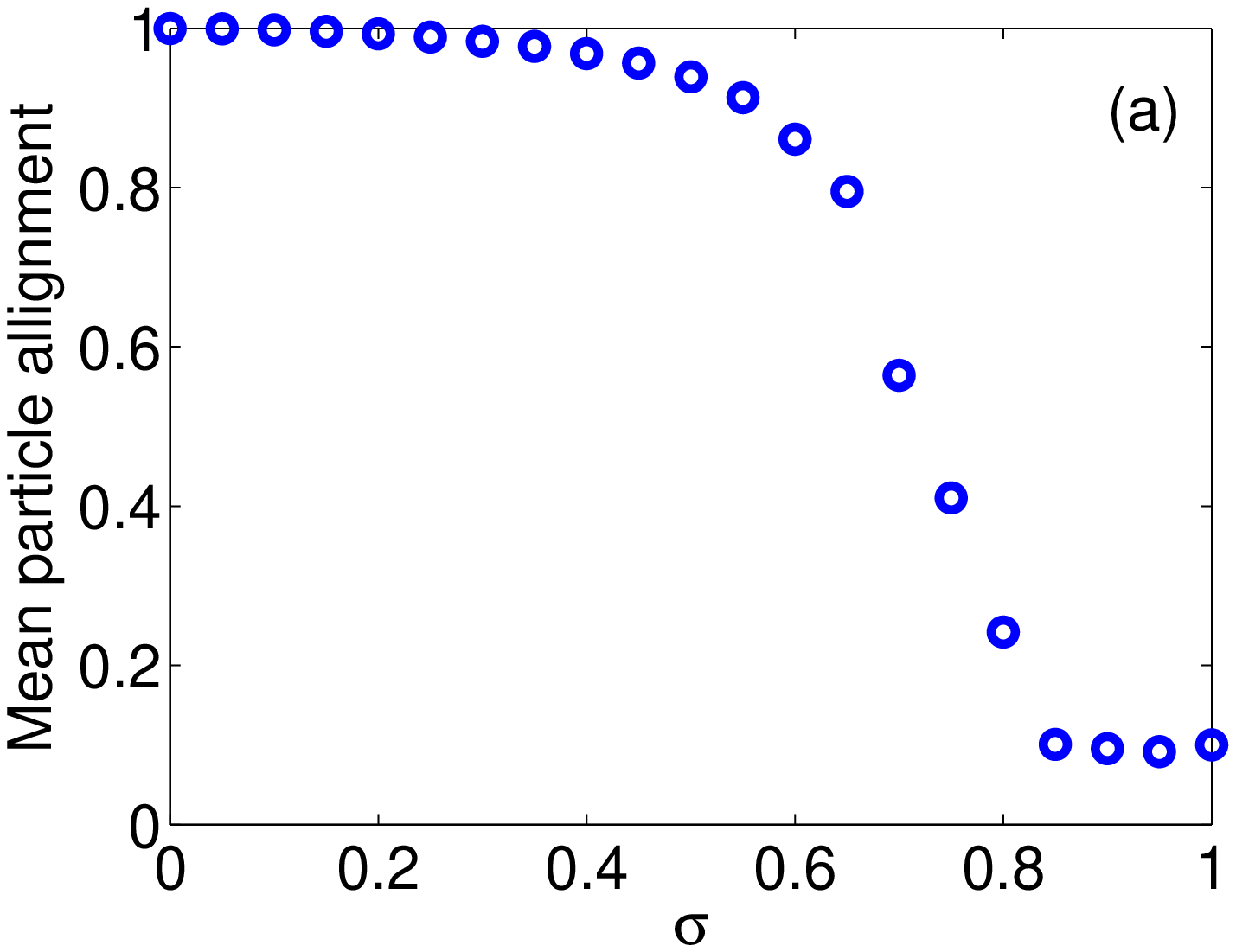} \label{mean_allign_a}}
\subfigure{\includegraphics[scale=0.26]{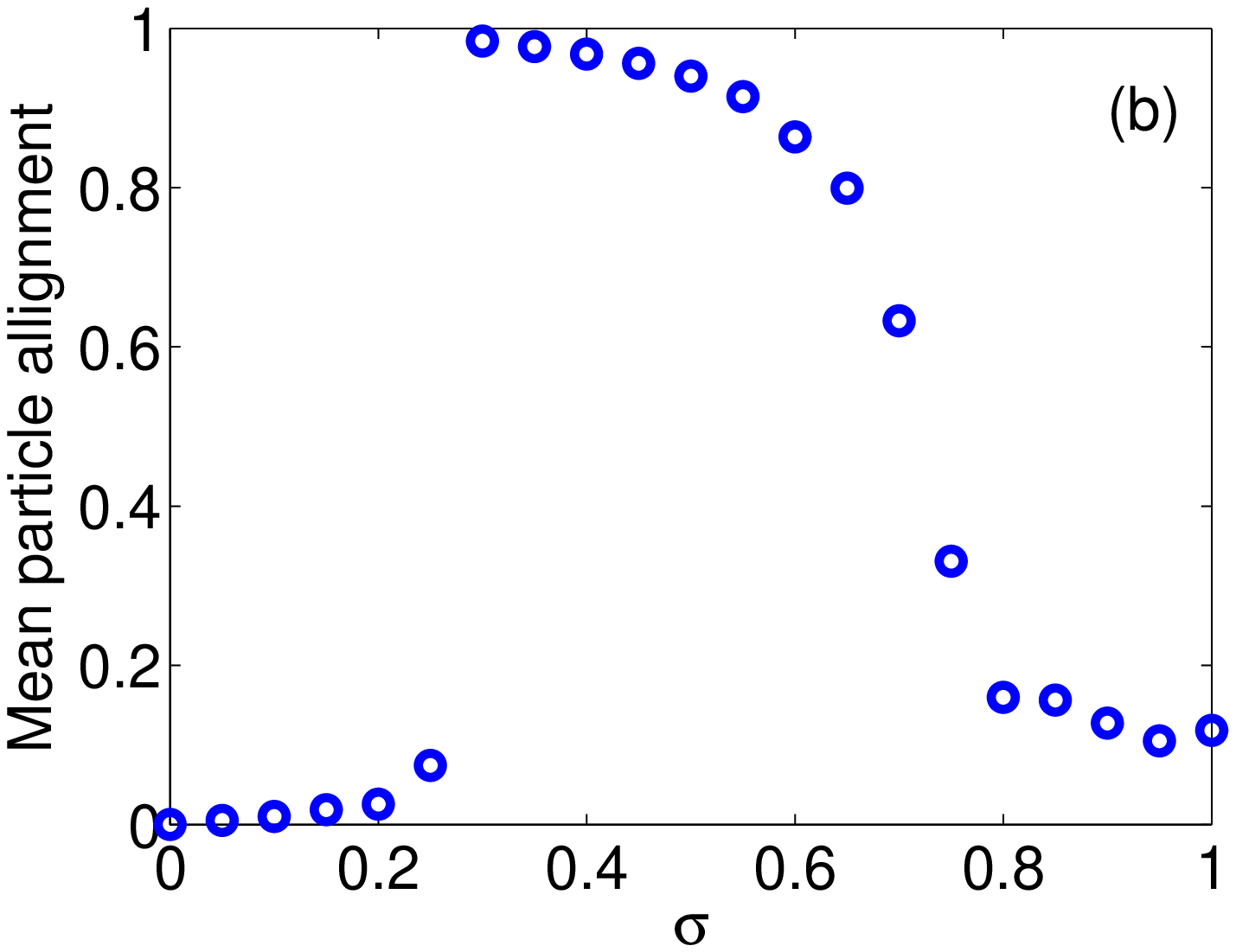} \label{mean_allign_b}}
\caption{Long time behavior of the mean particle alignment (defined in the text) for different
  values of noise intensity ($D=\sigma^2/2$) and two different initial
  conditions. In panel (a), all particles start off from the origin with
  equal velocity vectors; in panel (b), all particles start from rest,
  distributed uniformly over the unit square. For
  these simulations, $N=150$, $a = 2$ and $\tau = 2$. The time-delay is turned
  on at $t=50$, and the simulations run until $t=300$.}

\end{center}
\end{figure}

\addtolength{\textheight}{-1.5cm}

The presence of noise introduces interesting switching behavior that make the
initial conditions of the swarm critical in determining the long time behavior of the system. To demonstrate
this, we  have performed a series of simulations for different noise intensities
and two different initial conditions: (i) all particles start at the origin
with unit $x$ and $y$ speeds [Fig. \ref{mean_allign_a}] and
(ii) all particles are distributed uniformly over the unit square  and start
from rest [Fig. \ref{mean_allign_b}]. The simulations are run until $t=300$
using a coupling constant $a=2$ and a time-delay $\tau = 2$ which is
turned on at $t=50$. Our simulations reveal that in the long time limit and for small values of noise, the swarm converges to
either a compact state that rotates as a whole [case (i)] or to a ring
state with particles going both clockwise and counterclockwise [case
(ii)]. The asymptotic behavior of the system is readily visualized by calculating
the mean alignment of the swarm particles. We quantify this mean alignment of the swarm by calculating the
cosine between the directions of the $i$-th particle and the center of mass,
$\cos\theta_i=(\dot{\mathbf{ r}}_i\cdot\dot{\mathbf{R}})/(|\dot{\mathbf{r}}_i| |\dot{\mathbf{R}}|)$, and then averaging over all particles and over
the last 100 time units of simulation. Figure \ref{mean_allign_a} shows that in case (i) the particles
converge to the compact, aligned state for low and moderate noise intensities. However, this state is broken up at high noise levels ($\sigma
\approx 0.8$). In contrast, Fig. \ref{mean_allign_b} shows that in case
(ii) the particles converge to a ring for small values of noise ($\sigma
\lesssim 0.25$),
evidenced by the low values of the mean particle alignment in Fig. \ref{mean_allign_b}, but converge to the
aligned case for higher values of noise ($\sigma
\gtrsim 0.25$). Observing the full simulation runs in detail (not shown) reveals a switching
behavior: for case (ii) with a noise level $\sigma
\gtrsim 0.25$, the particles first converge to a noisy ring and then switch to the
rotating state due to the effect of noise. The simulations suggest that the
transition to the rotating state occurs once the velocities of the particles
cross an alignment threshold. The system, in fact, displays hysteresis: one can force the swarm
to transition from the ring state with $\sigma= 0.2$ to the rotating state
 by raising the noise to $\sigma= 0.25$; however, it seems that the inverse
 transition, i.e. making the
 swarm transition back to the ring state by lowering the noise level, is
 extremely unlikely.



\section{CONCLUSIONS}
To summarize, we studied the dynamics of a self-propelling swarm in the
presence of noise and a constant communication time delay and prove that
the delay induces a transition that depends upon the size of the interaction
coupling coefficient.  Although our analytical and numerical results were
obtained using a model with linear, attractive interactions, the analysis may
be applied to models with more general forms of social interaction.

We further uncovered a complete analytical description of the bifurcation
point which control the instabilities arising from noise induced
transitions. The analysis allows us to completely classify, using mean field
approximations, where the swarm exhibits a stable translation, stationary
center of mass, or rotation. 

In general, our results provide insight into the stability of complex systems comprised of
individuals interacting with one another with a finite time delay in a noisy
environment.  Furthermore, the results may prove to be useful in controlling
man-made vehicles where actuation and communication are delayed, as well as in
understanding swarm alignment in biological systems.

\section{ACKNOWLEDGMENTS}

The authors gratefully acknowledge the Office of Naval Research for their
support. LMR and IBS are supported by Award Number R01GM090204 from the
National Institute Of General Medical Sciences. The content is solely the
responsibility of the authors and does not necessarily represent the official
views of the National Institute Of General Medical Sciences or the National
Institutes of Health.  E.F. is
supported by the Naval Research Laboratory (Award No. N0017310-2-C007).


%
%
%
%
%


\end{document}